\documentclass[aps,prb,twocolumn,showpacs,superscriptaddress,floatfix]{revtex4-1}  % for review and submission
\usepackage{graphicx}  % needed for figures
\usepackage{dcolumn}   % needed for some tables
\usepackage{bm}        % for math
\usepackage{amssymb}   % for math
\usepackage{multirow}
\usepackage{amsmath}
\usepackage{natbib}
\newcommand{\affA}{CNRS-Laboratoire de Photonique et de Nanostructures, route de Nozay, F-91460, Marcoussis, France}
\newcommand{\affB}{Laboratoire de Physico-chimie des Microstructures et Micro-syst\`emes, Institut Pr\'eparatoire aux Etudes Scientifiques et Techniques, BP51, 2070 La Marsa, Tunisia}
\newcommand{\affC}{FOTON, Universit\'e Europ\'eenne de Bretagne, INSA-Rennes and CNRS, Rennes, France}

\begin{document}

\title{Microscopic Electronic Wavefunction and interactions between quasi particles in Empirical Tight-Binding Theory}

\author{R.~Benchamekh}\affiliation{\affA}\affiliation{\affB}
\author{F.~Raouafi}
\affiliation{\affB}
\author{J.~Even}
\affiliation{\affC}
\author{F.~Ben Cheikh Larbi}
\affiliation{\affB}
\author{P.~Voisin}
\affiliation{\affA} 
\author{J-M.~Jancu}
\affiliation{\affC}

\date{\today}
\begin{abstract}

A procedure to obtain single-electron wavefunctions within the tight-binding formalism is proposed. It is based on linear combinations of Slater-type orbitals  whose screening coefficients are extracted from the optical matrix elements of the tight-binding Hamiltonian. Bloch functions obtained for zinc-blende semiconductors in the extended-basis spds* tight-binding model demonstrate very good agreement with first-principles wavefunctions. We apply this method to the calculation of electron-hole exchange interaction, and obtain the dispersion of excitonic fine structure of bulk GaAs. Beyond semiconductor nanostructures, this work is a fundamental step toward modeling many-body effects from post-processing single particle wavefunctions within the tight-binding theory.
\end{abstract}

\maketitle

\section{Introduction}

Tight-binding is widely used as a conceptual frame to account for the kinetic energy operator in solid state theory. It served as a basis for major contributions such as Anderson's strong localization\cite{Anderson1} and magnetic impurity \cite{Anderson2} theories, Hubbard's model of interacting electrons\cite{Hubbard}, and many others\cite{Wallace,Fu,Saito}. In these theories, interactions are generally introduced using symmetry considerations and ad-hoc parameters, and the goal is to have the simplest model integrating the physically relevant features of hopping and interaction integrals while getting rid of the complexities of underlying atomic physics.
On the other hand, empirical-parameter tight-binding (EPTB) is known as a powerful modeling technique for the electronic structure of semiconductors, metals, and all kind of nanoscale structures and devices. The systematic procedure for constructing the EPTB Hamiltonian was discussed in a seminal paper by Slater and Koster (SK) in 1954 \cite{slater54}, but it is only many years later that computers have allowed systematic implementations \cite{Vogl, Schulman}. A major step was achieved in the late 1990's with the development of EPTB models using an extended spds* orbital basis \cite{Jancu98, Barreteau} and allowing accurate full-band representation of single particle states. However, when it comes to calculating short-range interactions between quasi-particles, EPTB models (like the $\bf{k.p} $ theory itself) are hampered by the lack of an explicit orbital basis. This is a strong limitation for the use of advanced EPTB schemes to explore highly topical problems of strongly correlated electron systems. In this paper, we present a method that reconciles the ``conceptual frame'' and ``modeling tool'' faces of tight-binding, by self-consistently determining the orbital basis out of the EPTB hamiltonian. We thus obtain the local wavefunctions, which allows parameter-free calculation of short-range interactions. We illustrate this by calculating electron-hole exchange and the fine structure of excitons in GaAs. Beyond semiconductors, this method can be used for many different materials and can handle million-atom supercells that are still out of reach of first-principle methods.

\section{self-consistent projection basis for the EPTB Hamiltonian}

In the SK formalism, the crystal potential is approximated as a sum of spherically symmetrical potentials around each atom. This allows the electronic wavefunctions $\Psi_{\alpha\mathbf{k}} $, where $\alpha$ stands for the band indices
to be developed on a set of Bloch sums $\Phi_{lm\mathbf{k}}$ of atomic-like orbitals (called the L\"owdin orbitals) $\phi_{lmj} $, $\phi_{lmj} $ is  the $m^{th}$ orbital on $l^{th}$ atom in $j^{th}$ unit cell:
\begin{equation}\Psi_{\alpha\textbf{k}}= \sum_{m,l} C_{ml}^{\alpha} \Phi_{lm\textbf{k}} \label{eq:1.6}\end{equation}
\begin{equation}\Phi_{lm\textbf{k}}=\frac{1}{\sqrt{N}} \sum_{j} \exp (i \textbf{r}_{jl}. \textbf{k}) \phi_{ml}(\textbf{r}-\textbf{r}_{jl}) \label{eq:1.5}\end{equation}

The L\"owdin orbitals have well defined angular properties, but unknown radial dependencies. The hamiltonian matrix elements between them are treated as ``disposable constants'' with which one can fit band structures that have been experimentally determined or calculated using more accurate techniques. The fit is performed in k-space, removing  any necessity to further characterize the local wavefunctions in real space.
Very interestingly, following a method introduced by Boykin and Vogl, interaction with the electromagnetic field can be built-in using a derivation of the Hamiltonian matrix elements in momentum space\cite{boykin01}: 
\begin{equation}  \textbf{\textit{p}}_{lm,l'm'}= i \hbar \langle\Phi_{lm}\mid \bigtriangledown_{\textbf{\textit{k}}}H \mid\Phi_{l'm'}\rangle \end{equation}   

Optical properties can be consequently calculated in EPTB models without adding parameters, and (although this method misses intra-atomic matrix elements) good accuracy is obtained providing that the orbital basis is rich enough \cite{foreman}. In this context, the spds* TB model is known to give a description of dielectric properties equivalent to best ab initio calculations within the one-electron approximation \cite{TB1}. Altogether, the model major qualities are the transferability of parameters from bulk materials to nanostructures, the unique ability to describe with the same accuracy electronic properties in any region of the Brillouin zone, and the capacity to handle large supercells.
However, interactions (in particular, short range interactions) between quasi-particles involve the  local wavefunctions, and can not be calculated in this frame. Existing calculations of Coulomb matrix elements use approximations on the radial dependence of the basis orbitals \cite{schulz,Korkusi}. 
\par
In order to solve this theoretical issue that has remained open ever since the seminal work of Slater and Koster, we start with a trial set $\mathcal{B}$ of $spds^*$ basis functions in the form of normalized Slater-type orbitals (STO) 
$\phi^t_{nlm}(\textbf{\textit{r}})=\sqrt{{(2\alpha)^{2n+1}}/{(2n)!}}$ $ Y_{lm}(\theta,\phi)  r^{n-1} e^{-\alpha r} $, where $n$ is the first quantum number and $\alpha$ is a screening parameter\cite{slater30}. STOs are largely employed in quantum chemistry but do not fulfill the orthogonality condition, since finite overlap exists between two STOs localized at different sites of the crystal. Firstly, the orbital overlap matrix S is calculated including all orbitals up to a cut-off distance $R_0$ that must be taken large enough so that overlap with remote atoms be negligible. Thanks to the nice properties of STOs, this step is done analytically, and in practice, we found that for $\alpha=0.5$, overlap with neighbors located farther than 3 lattice parameters ($17$\AA ) can safely be neglected. Note that, unlike real atomic orbitals, $s$ and $s*$ on-site STOs are not orthogonal. This small difficulty is easily solved by substituting $s^*$ with a Gramm-Schmitt combination $\Tilde{s}^*= (s^*- \langle s \shortmid s^*\rangle s)/\sqrt{1-{\langle s \shortmid s^*\rangle}^2}$. Then, an orthogonal basis $ \mathcal{B}_{orth} $ can be obtained using the L\"owdin orthogonalization procedure $ \mathcal{B}_{orth} =S^{-\frac{1}{2}}\mathcal{B}$ \cite{lowdin}. The orthogonalized STOs will serve as trial functions for the unknown L\"owdin orbitals. The expansion of electronic eigenfunctions (Bloch functions) in the basis $ \mathcal{B} $ is obtained by multiplying the eigenvectors of the $sp^3d^5s^*$ Hamiltonian matrix by the matrix $S^{\frac{1}{2}} $, which provides definitely their representation in real space.  Then the momentum matrix elements are calculated \textit{in real space} from the Bloch sums by the relation:
\begin{equation}  \textbf{\textit{p}}_{lm,l'm'}= i \hbar \langle\Phi_{lm}\mid \bigtriangledown_{\textbf{\textit{r}}} \mid\Phi_{l'm'}\rangle \end{equation} 
This derivation involves a sum of matrix elements between two STOs that are calculated analytically\cite{TB2}. Finally the screening parameters are fitted into a genetic algorithm until the optical matrix elements calculated in real space compare satisfactorily with those derived in k-space from the electronic Hamiltonian. In the end, the optical matrix element between two electronic bands denoted by $\alpha$ and $\beta$  is obtained by:
\begin{equation}
P_{\alpha,\beta}=\sum_{l,m}\sum_{l'm'} C_{lm}^{\alpha} \; \textbf{\textit{p}}_{lm,l'm'} \; C_{l'm'}^{\beta}
\end{equation}
In practice,any orthogonalization method could be used in this procedure. However, L\"owdin method has a unique merite that orthogonal basis set remains as close as possible to the original non-orthogonal set and retains their symmetry characteristics.

\begin{figure}
\includegraphics[width=40mm]{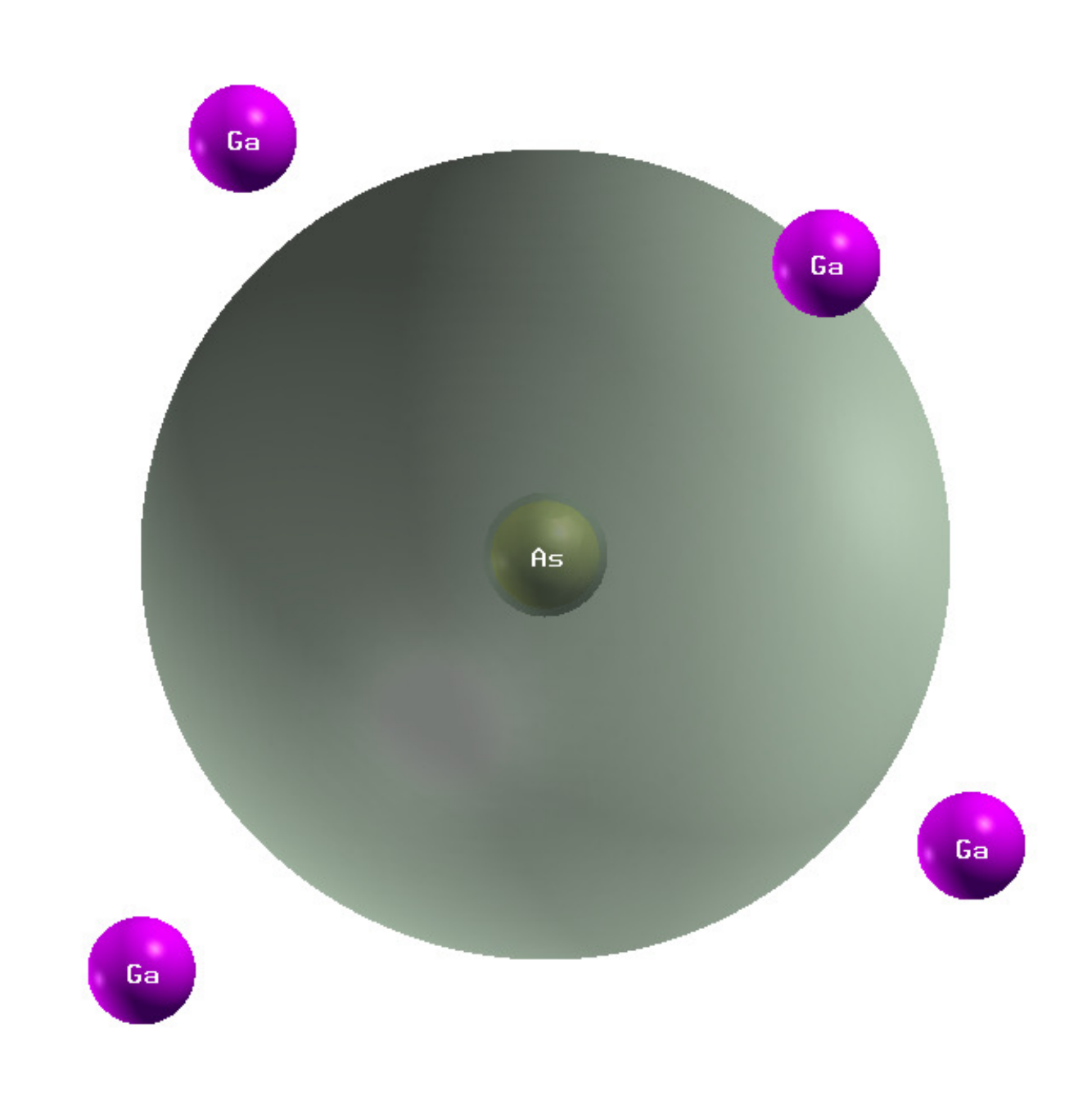}\includegraphics[width=40mm]{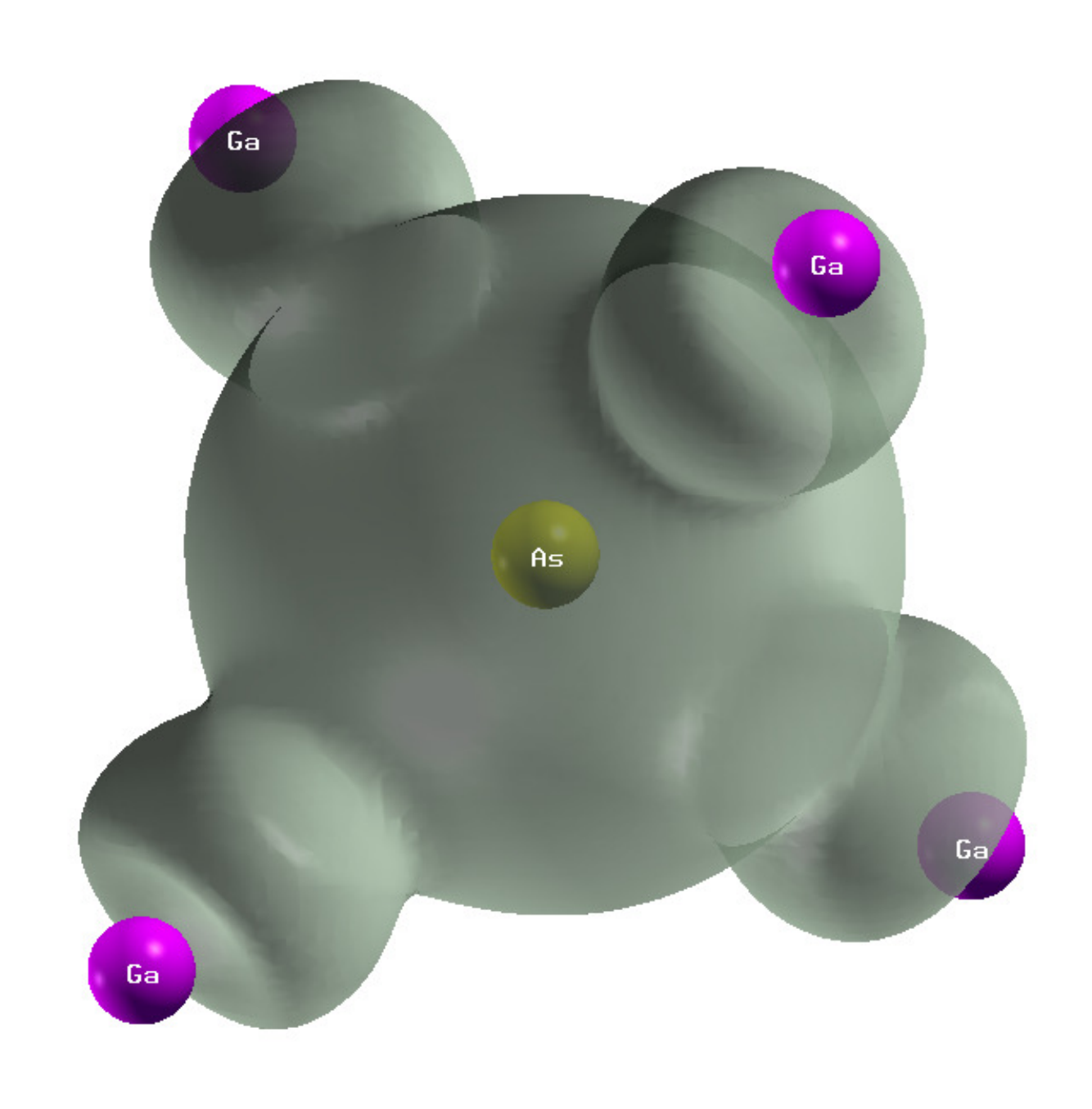}

\includegraphics[width=40mm]{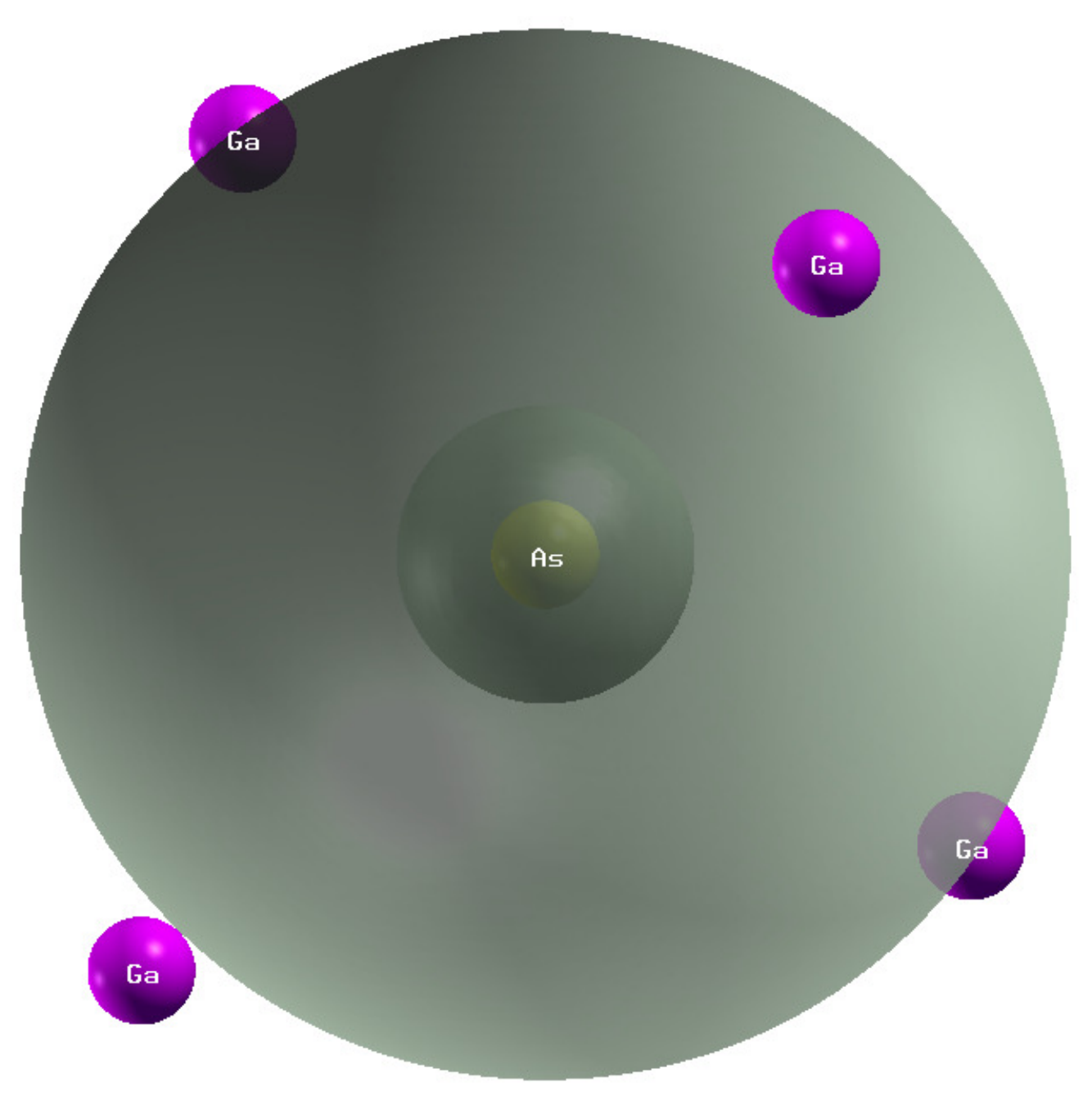}\includegraphics[width=40mm]{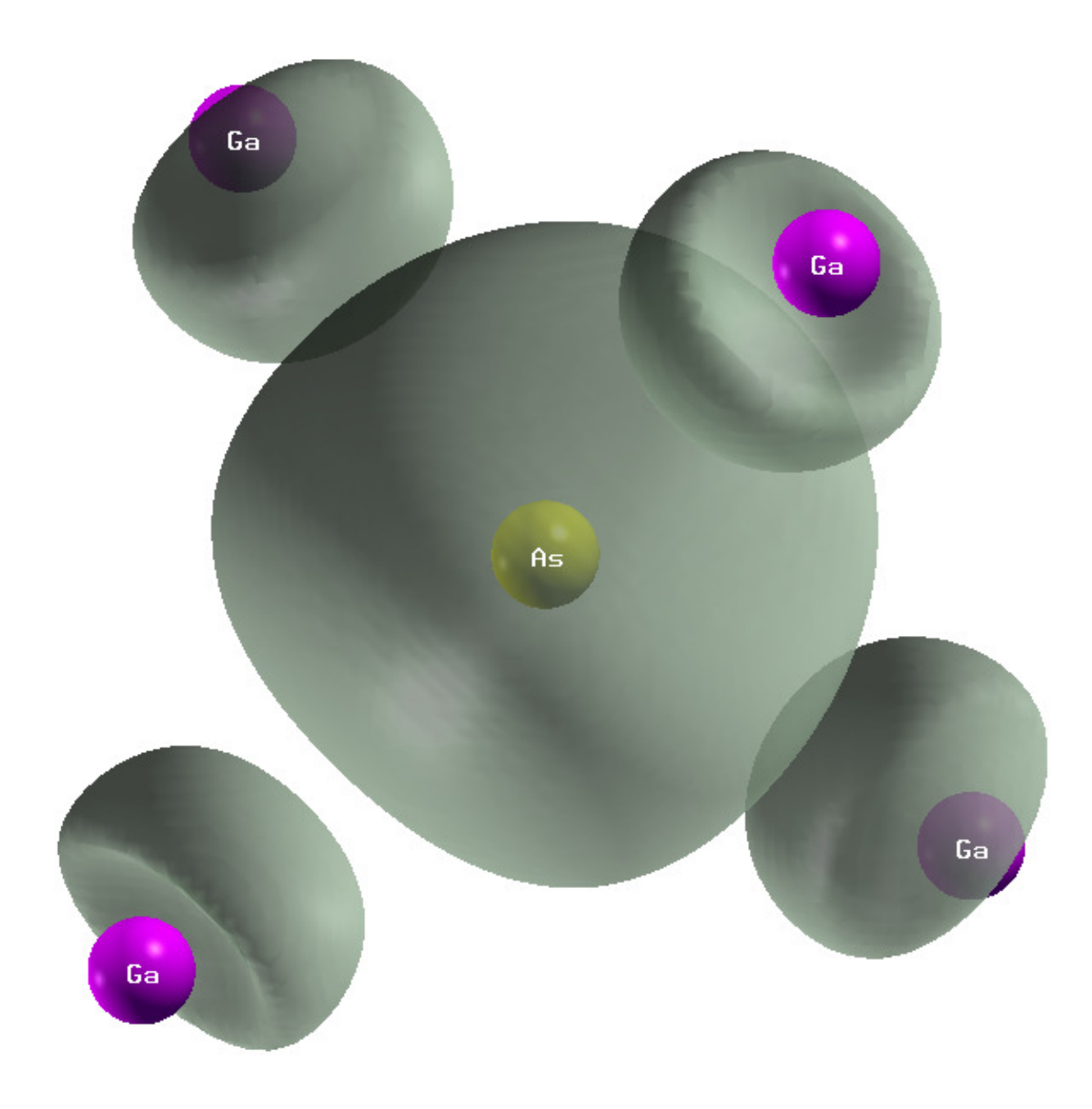}
\caption{Arsenic $s$ (top) and $s^*$ (bottom) Slater orbitals before (left) and after (right) orthogonalization. \label{fig:30}         }
\end{figure}

\begin{figure}
\includegraphics[width=40mm]{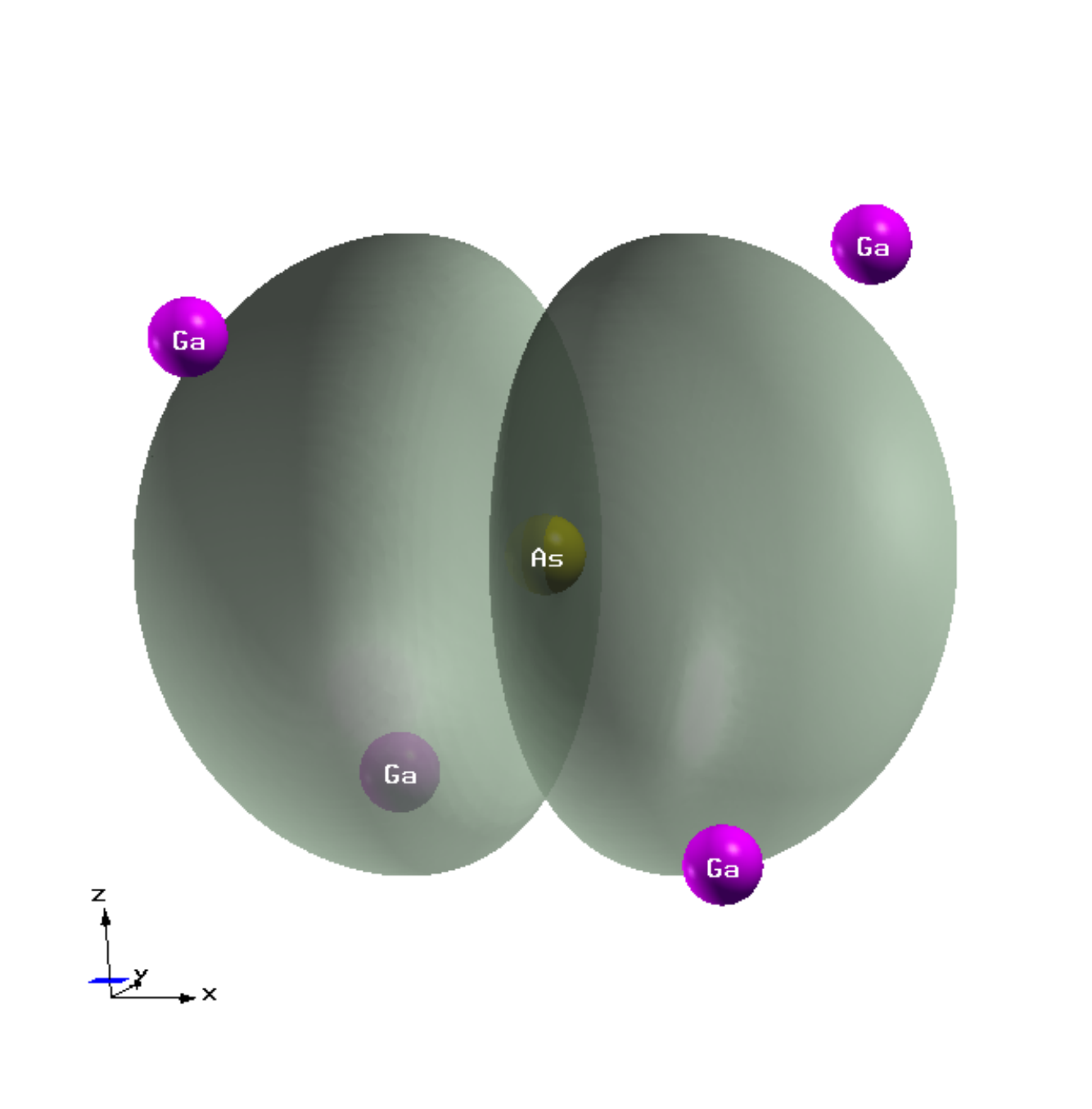}\includegraphics[width=40mm]{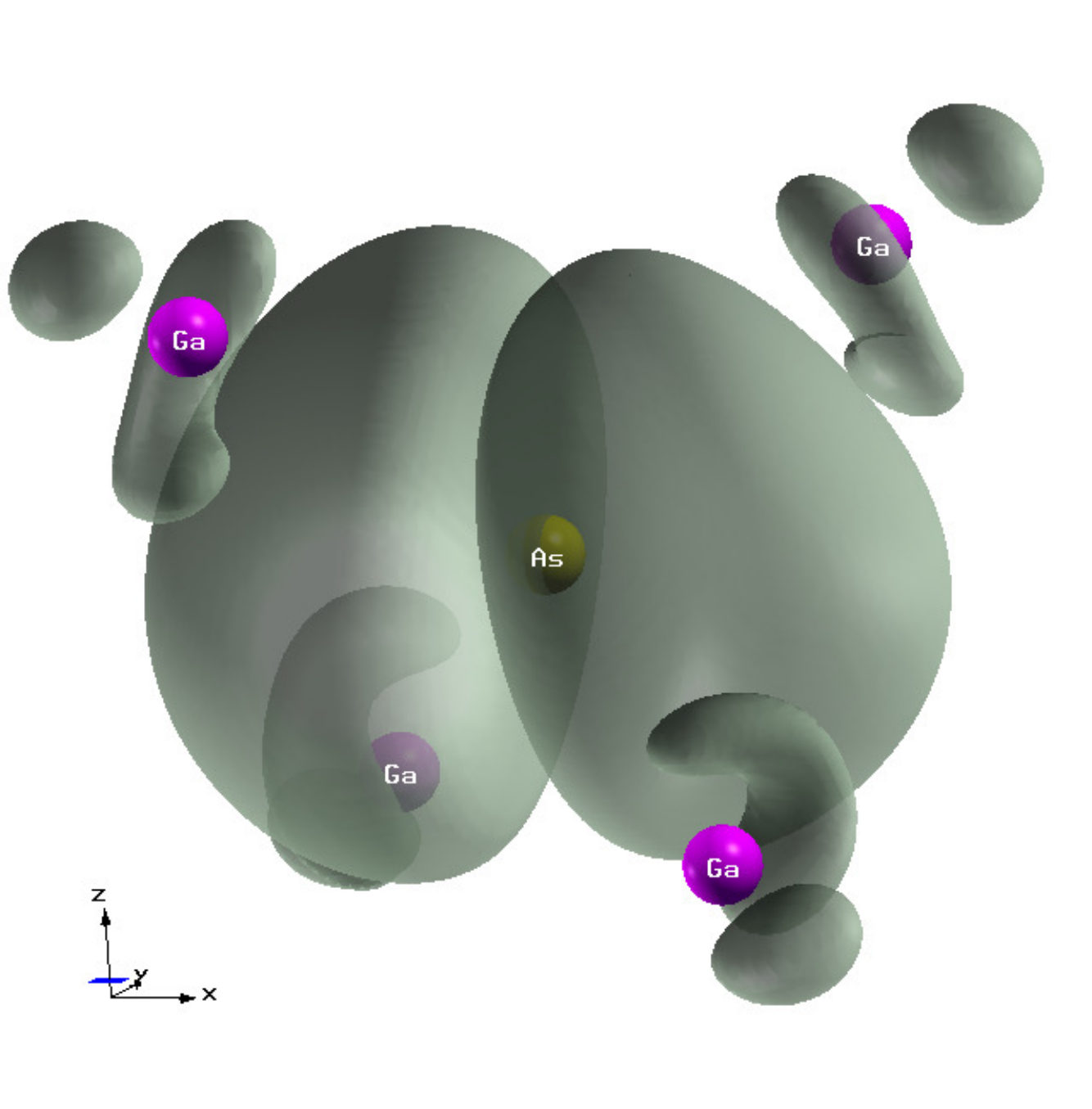}
\includegraphics[width=40mm]{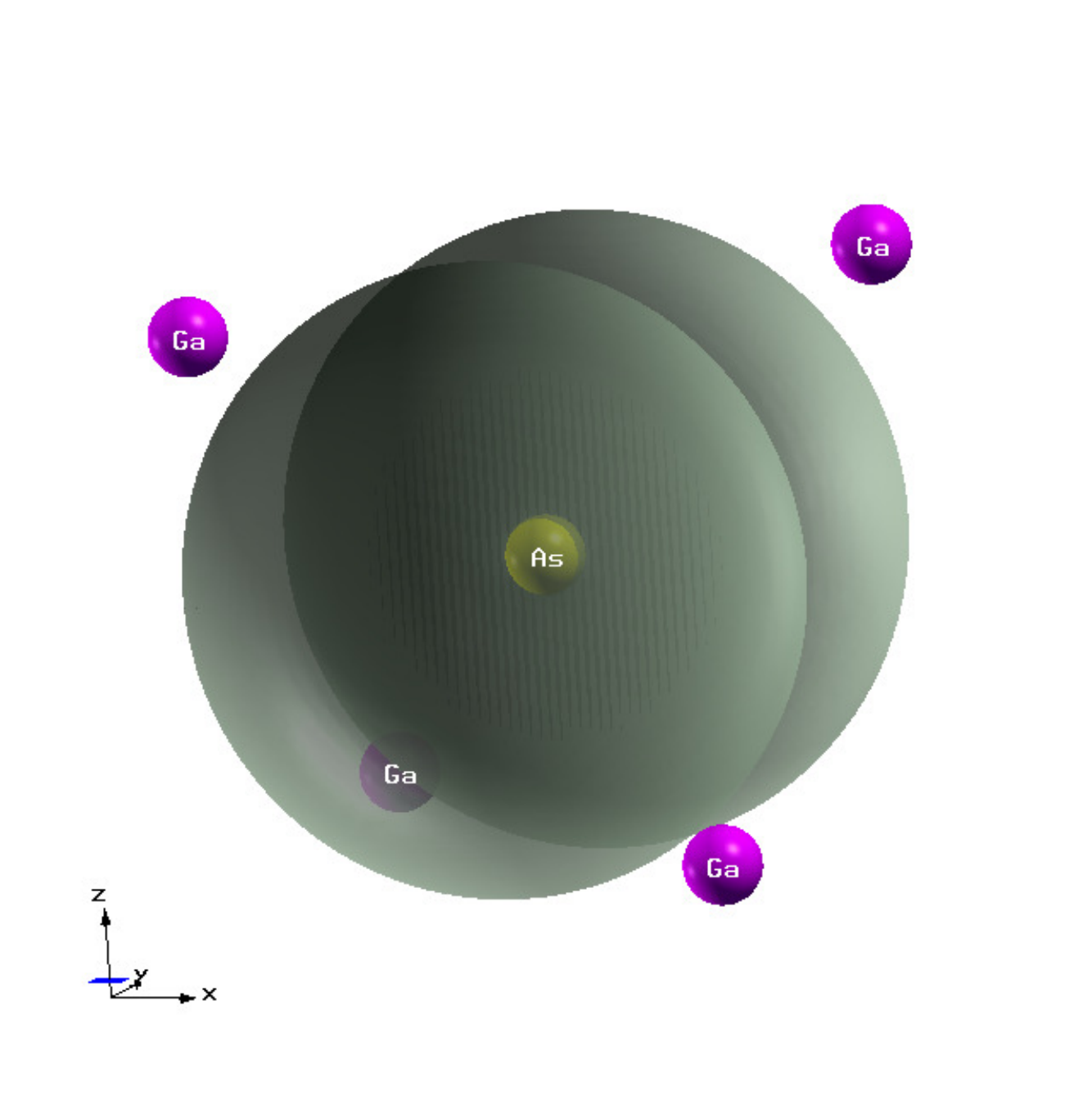}\includegraphics[width=40mm]{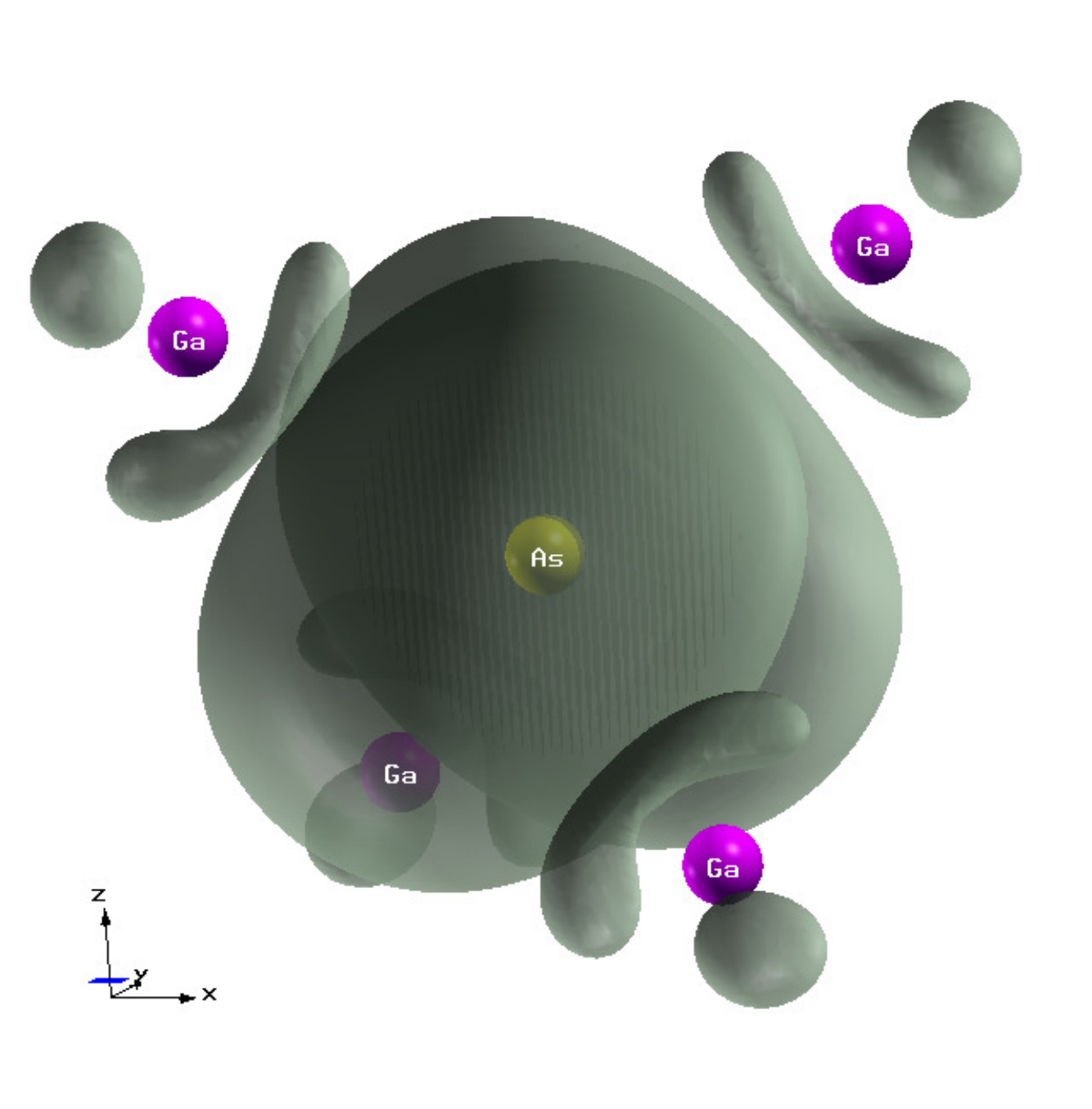}
\includegraphics[width=40mm]{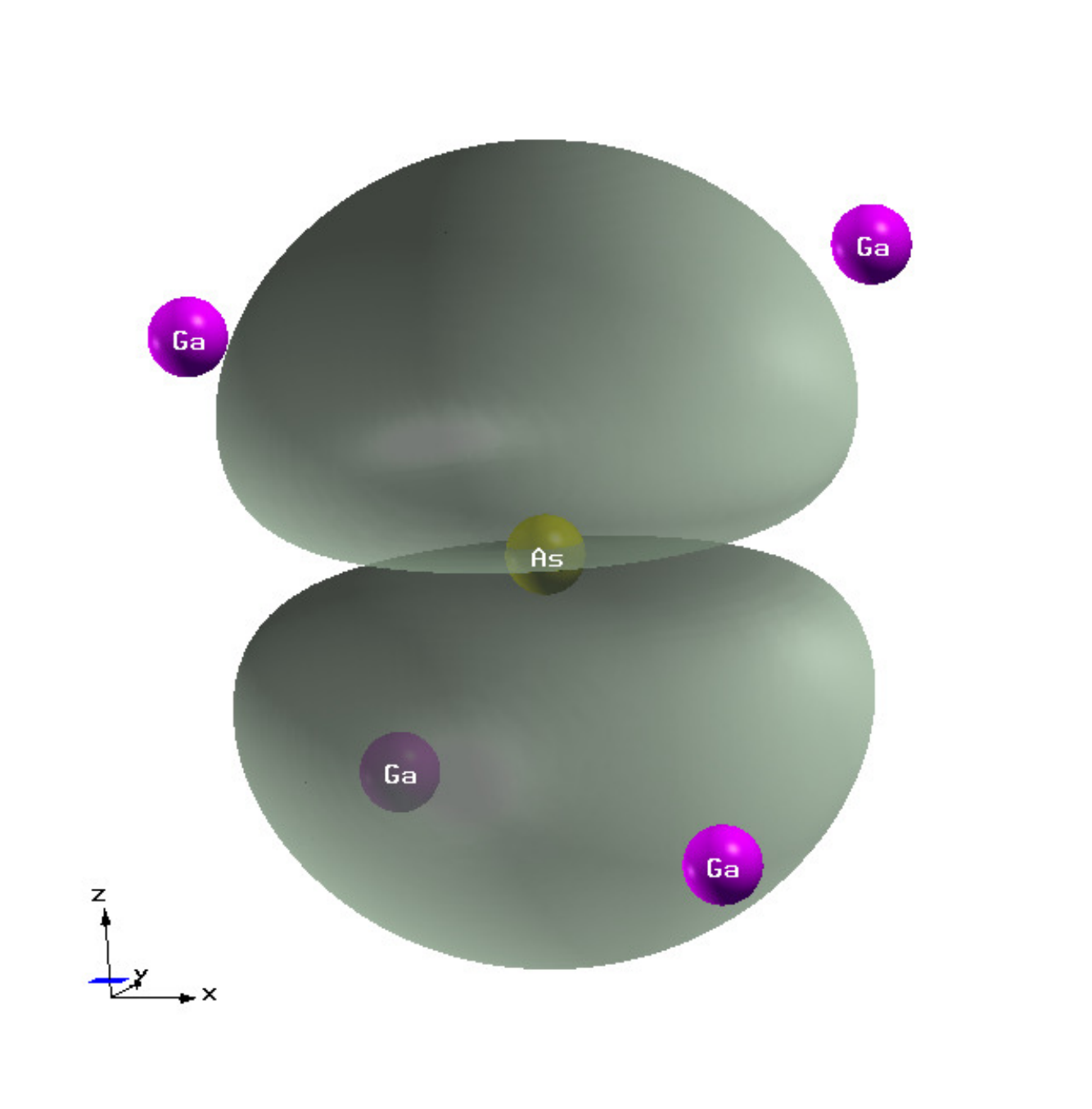}\includegraphics[width=40mm]{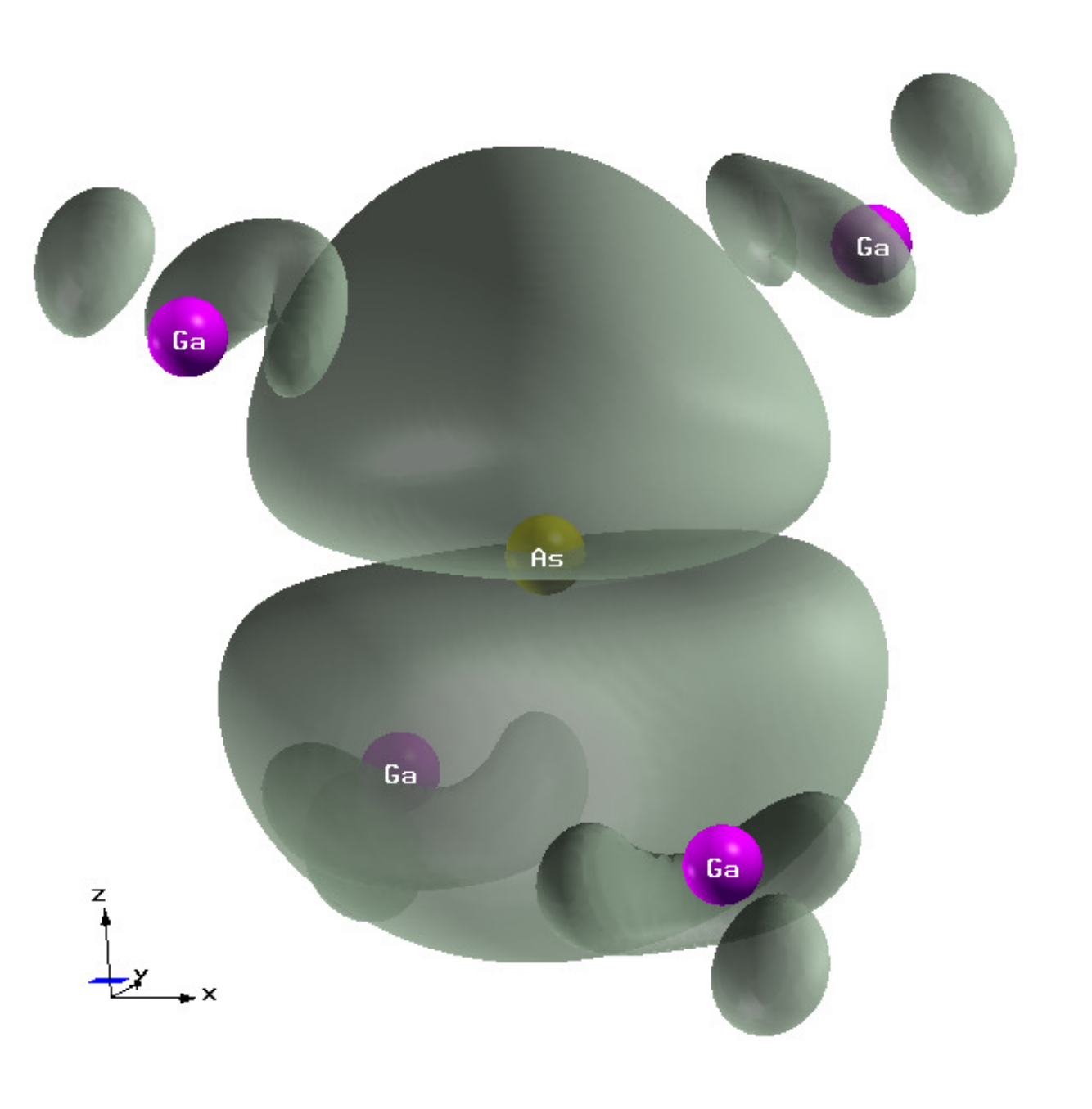}
\caption{Arsenic $p_x$,$p_y$ and $p_z$ Slater orbitals before (left) and after (right) orthogonalization.\label{fig:31}         }
\end{figure}

\begin{figure}
\includegraphics[width=40mm]{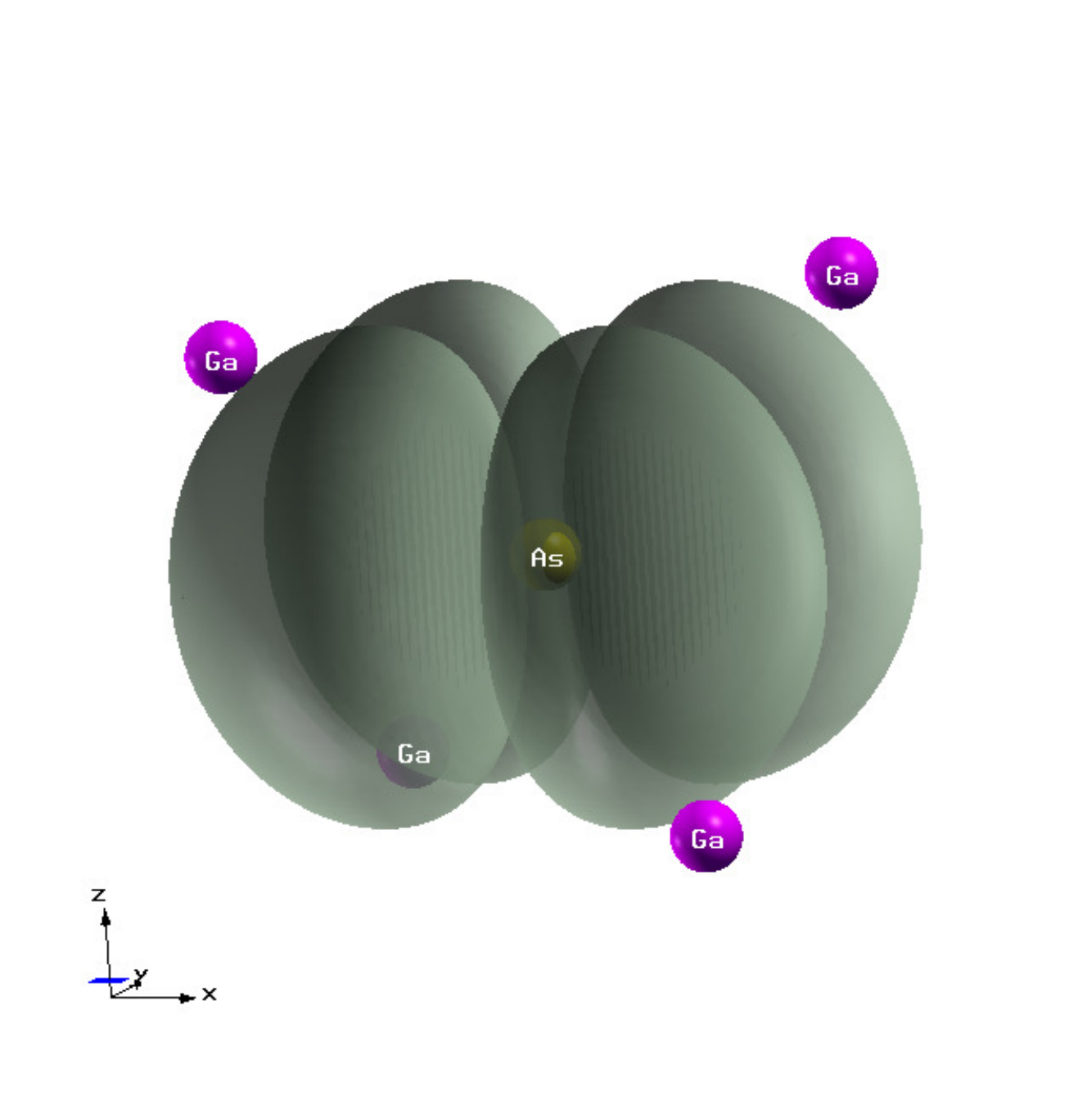}\includegraphics[width=40mm]{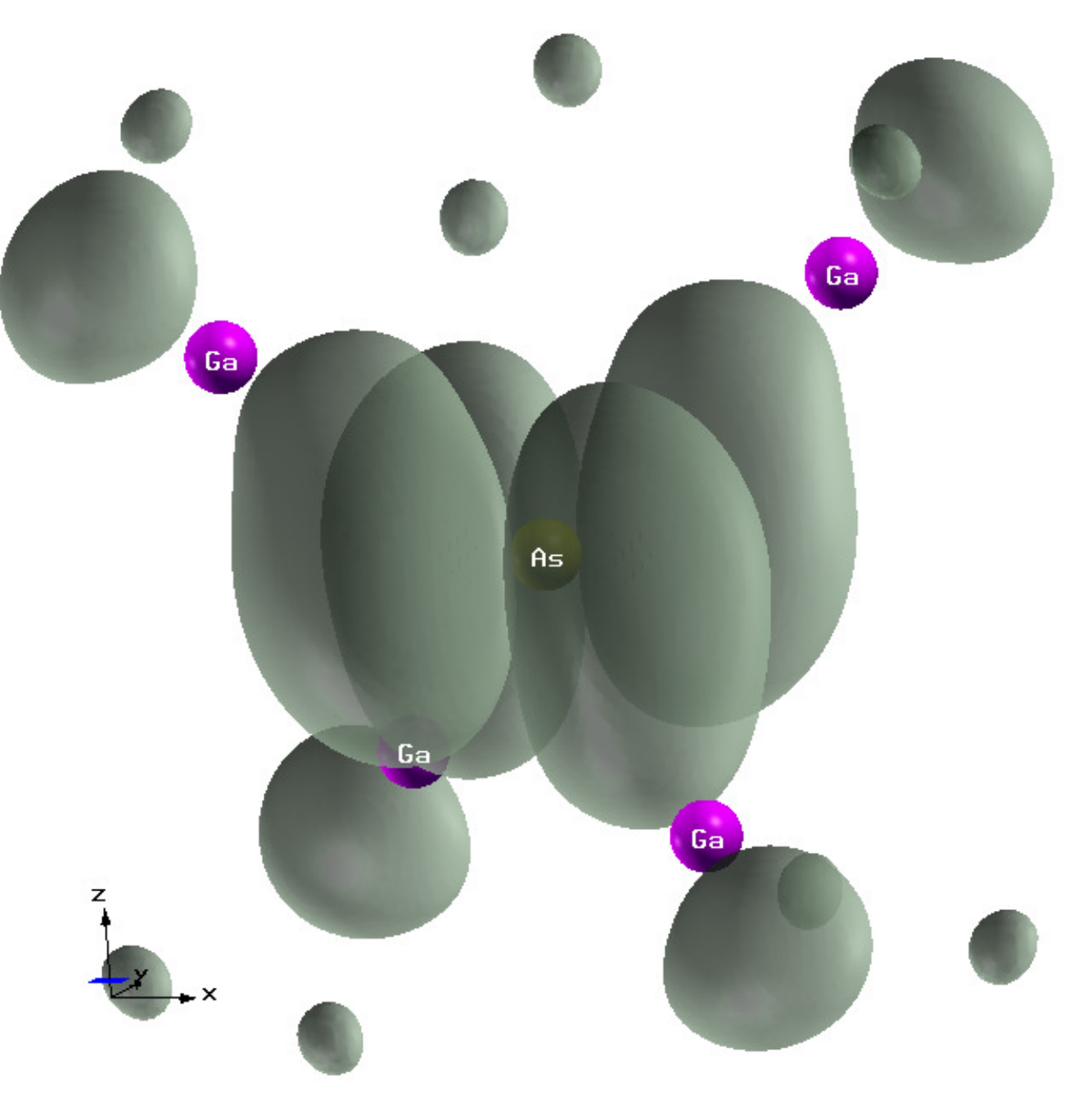}
\includegraphics[width=40mm]{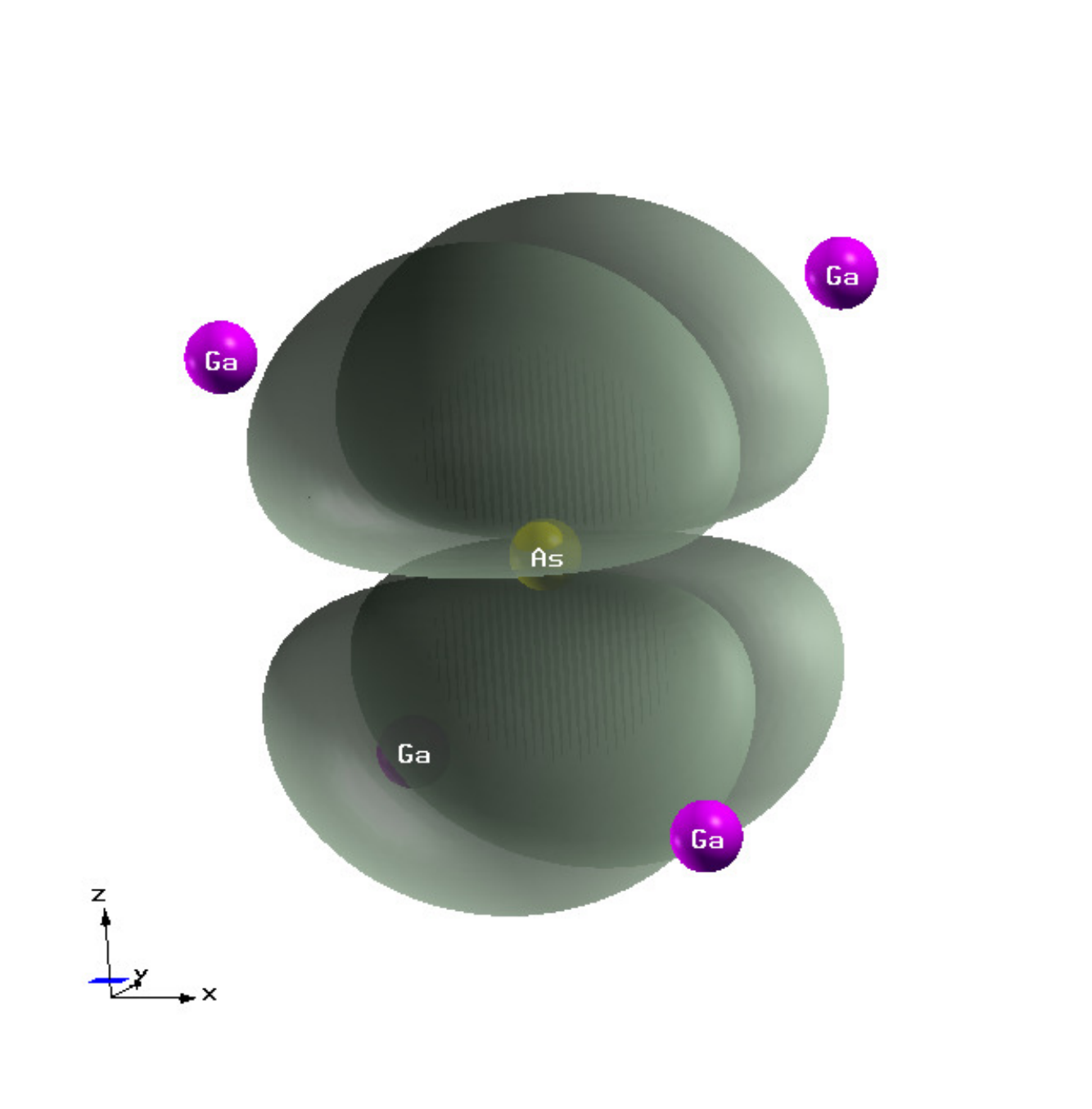}\includegraphics[width=40mm]{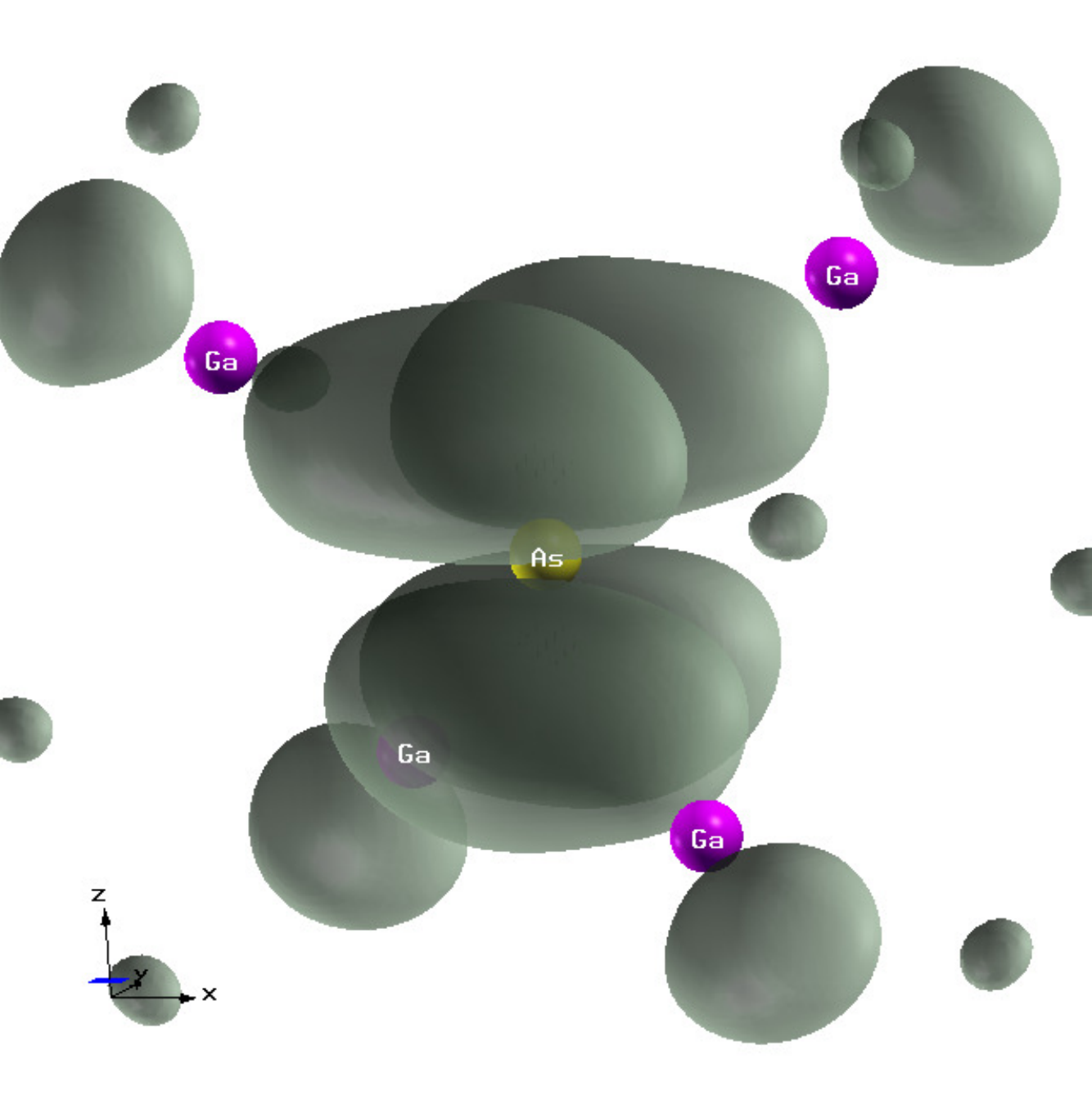}
\includegraphics[width=40mm]{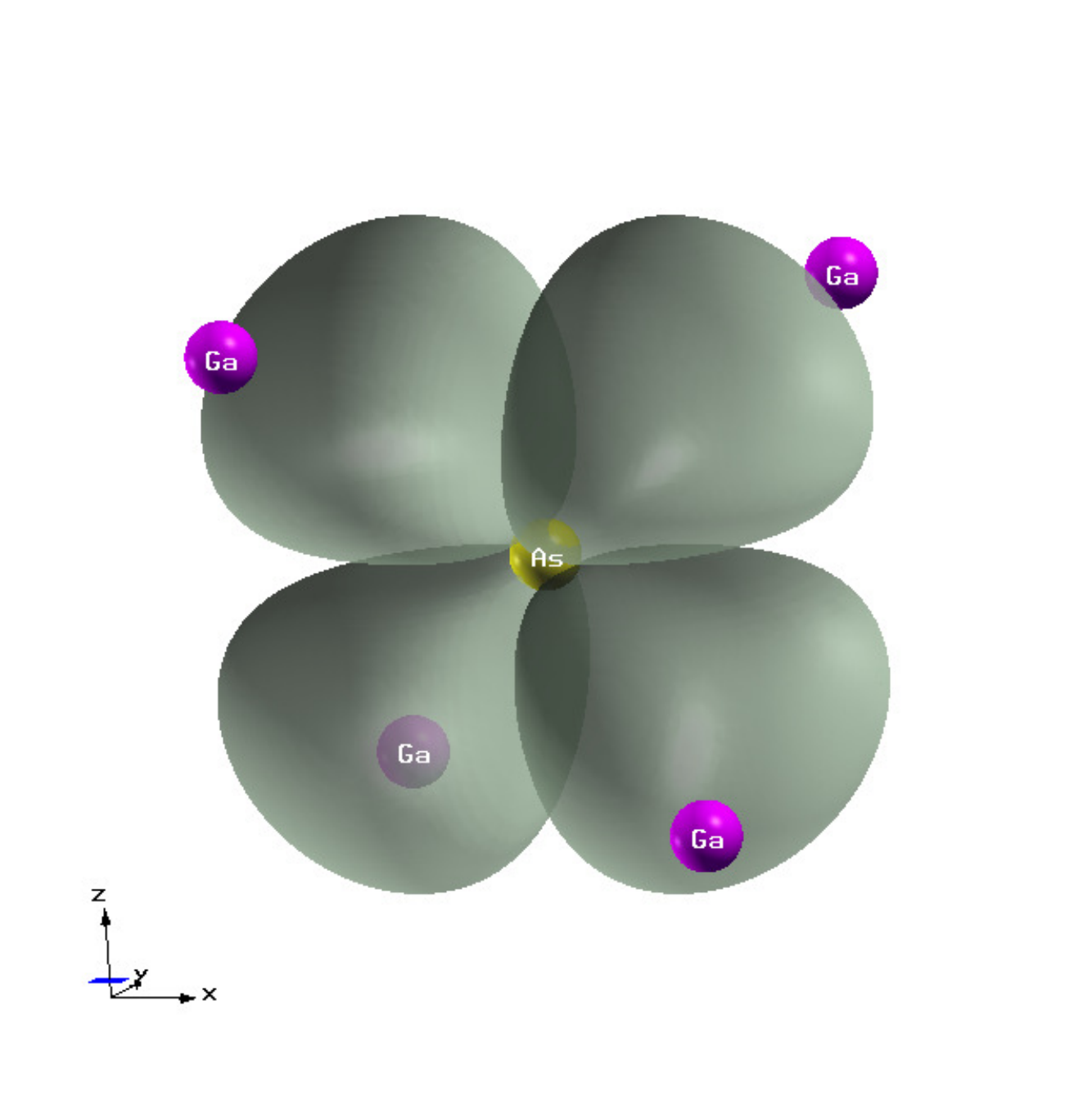}\includegraphics[width=40mm]{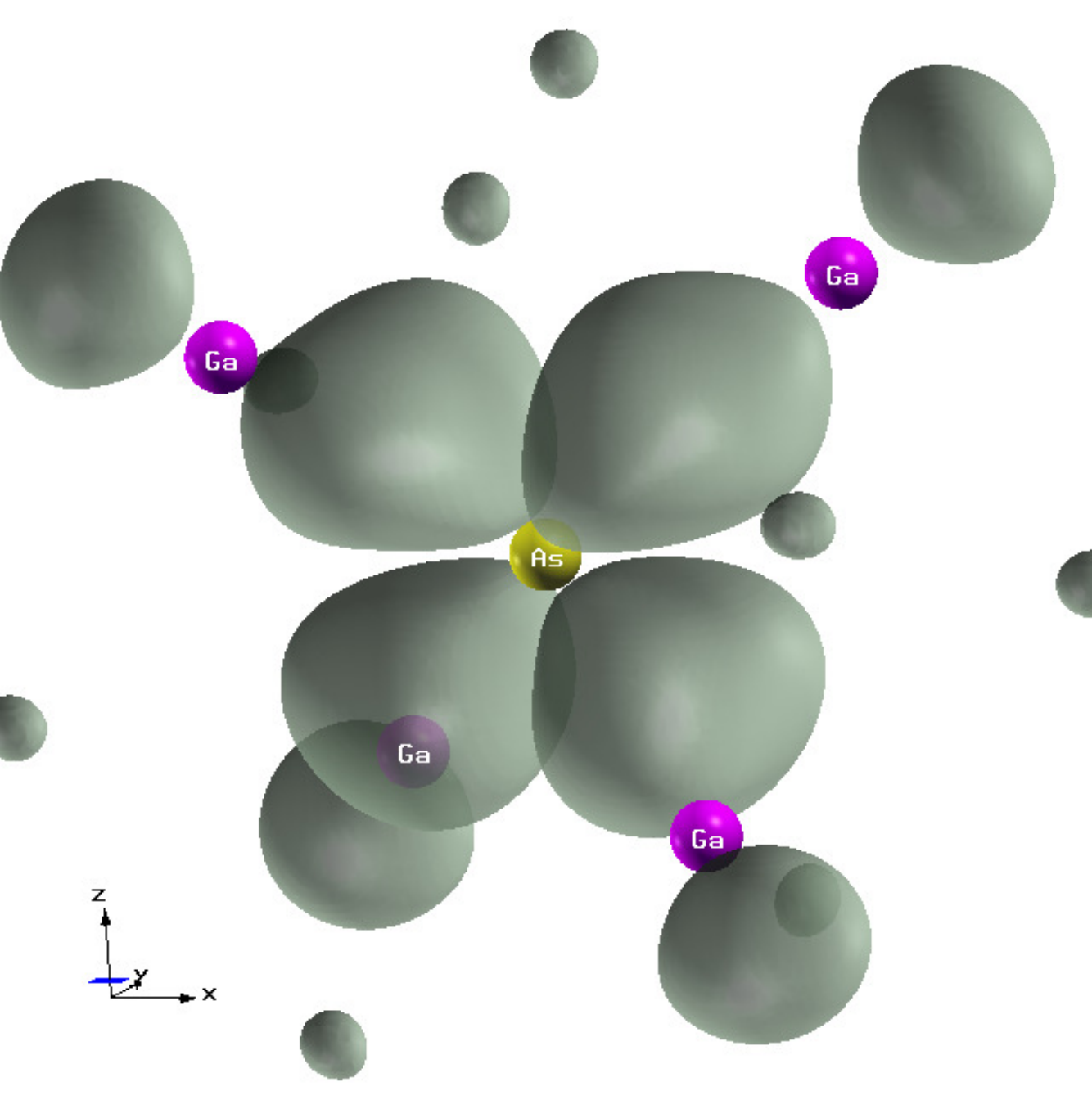}
\includegraphics[width=40mm]{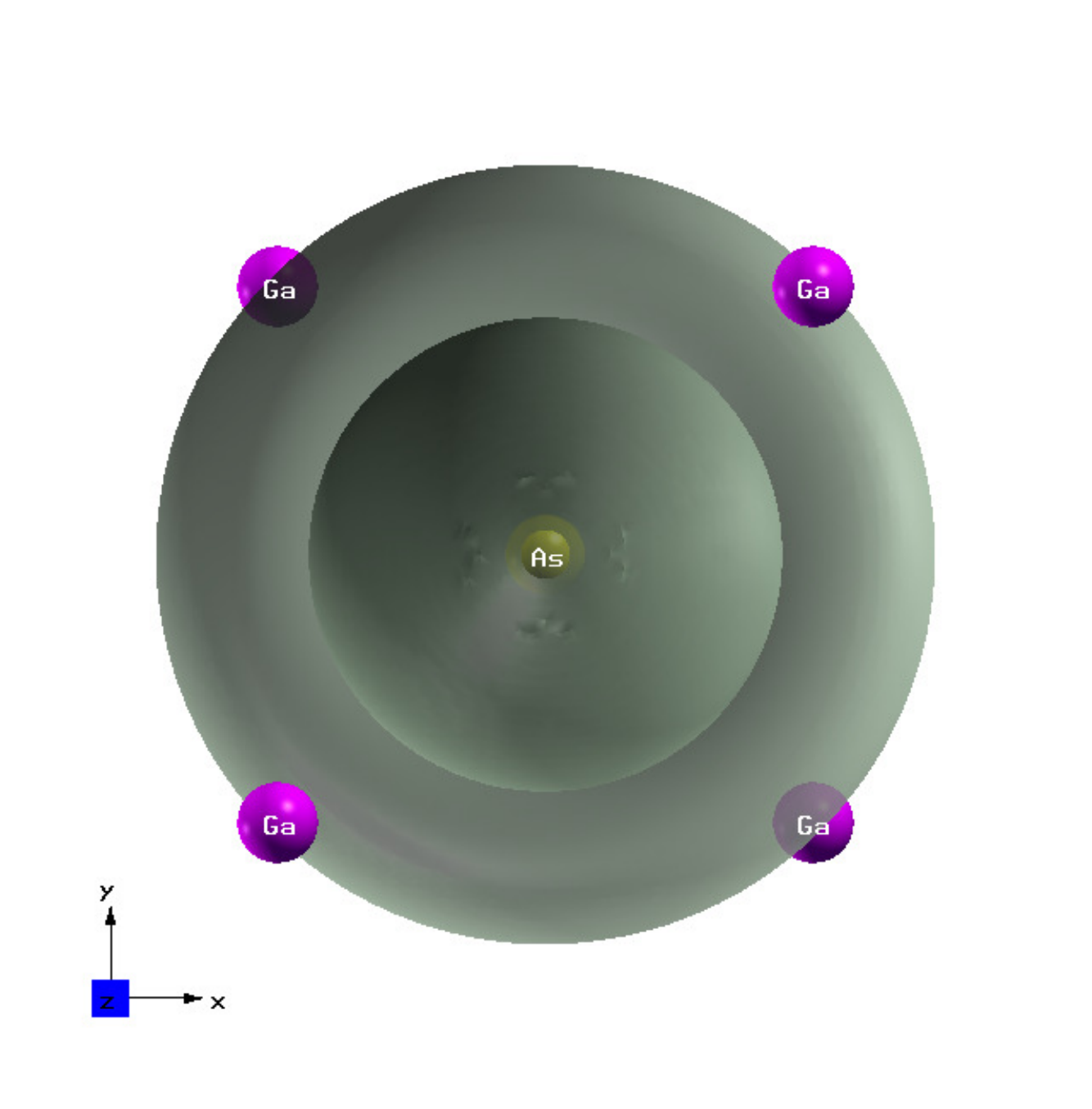}\includegraphics[width=40mm]{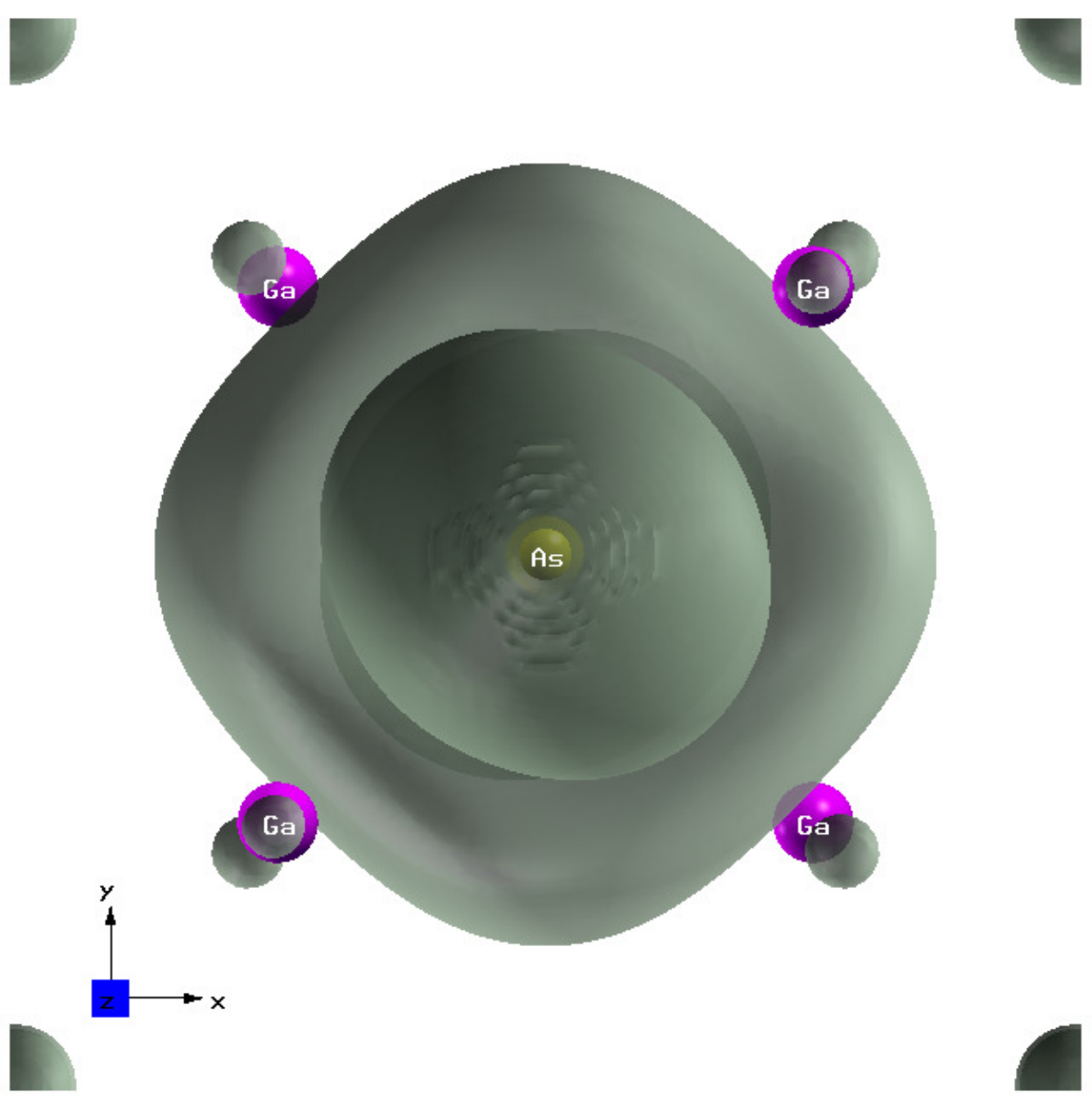}
\includegraphics[width=40mm]{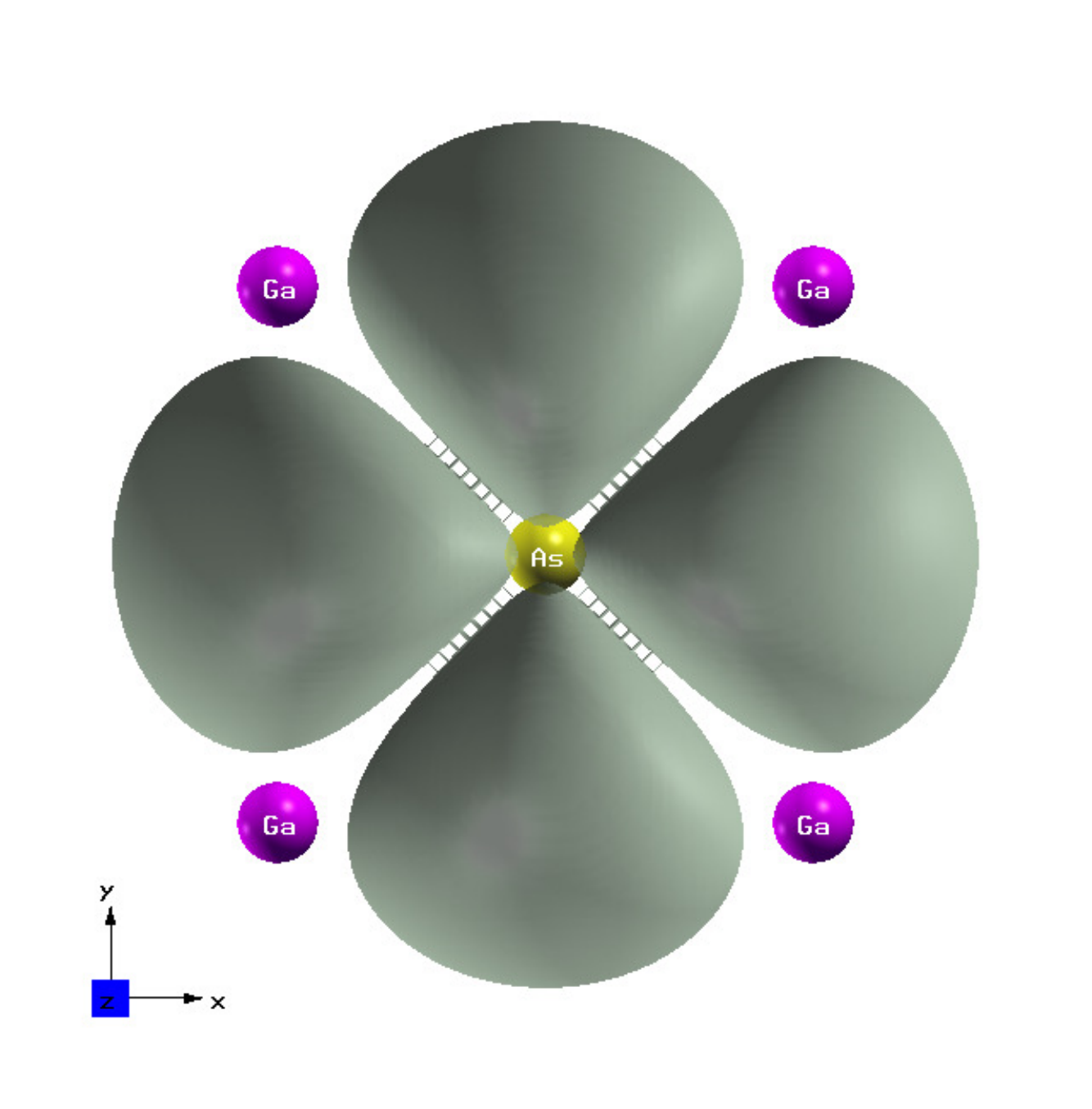}\includegraphics[width=40mm]{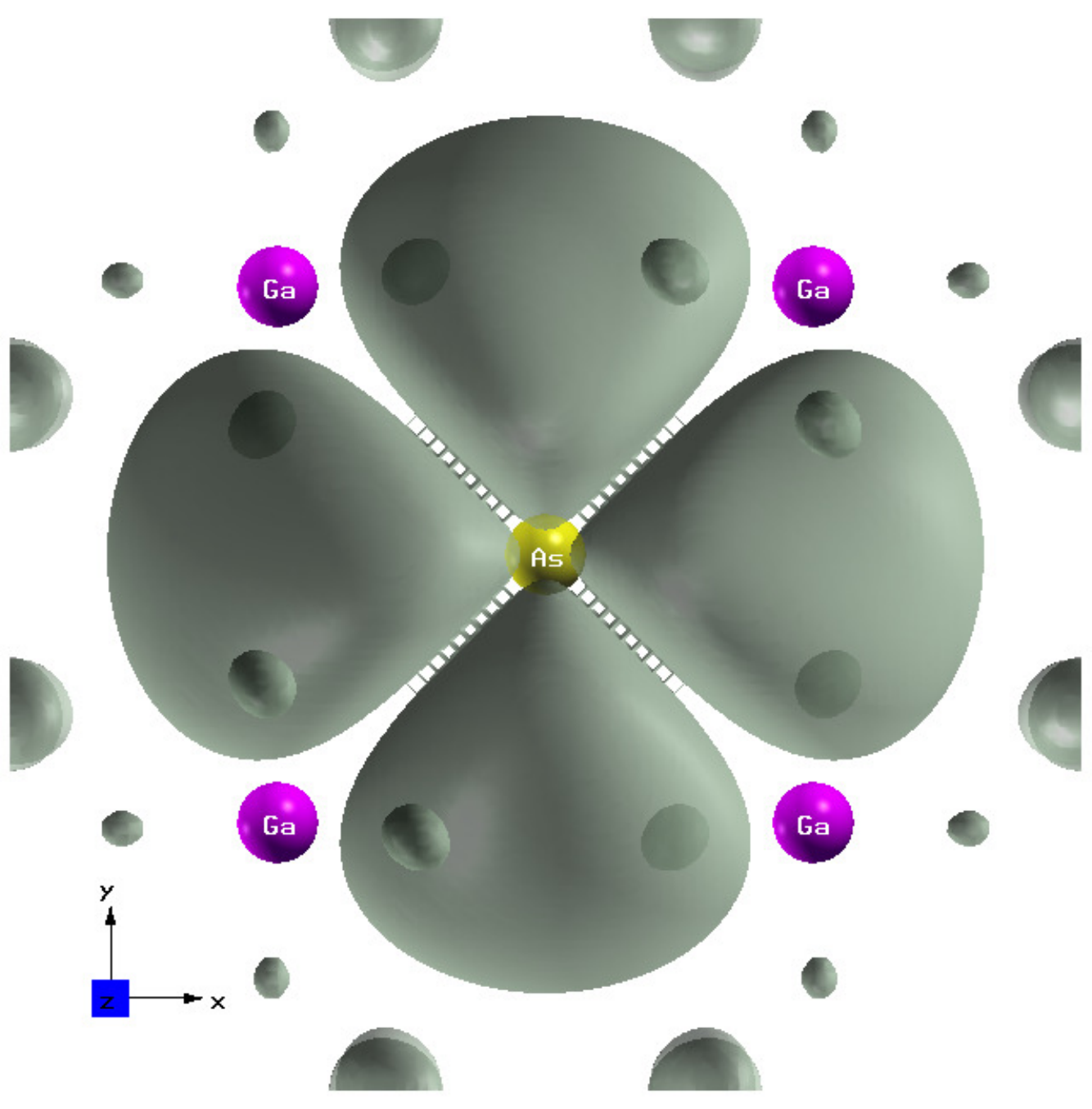}
\caption{Arsenic $d_{xy}$, $d_{yz}$, $p_{xz}$, $p_{3z^2-r^2}$ and $p_{x^2-y^2}$ Slater orbitals before (left) and after (right) orthogonalization.\label{fig:32}         }
\end{figure}

\section{Practical implementation and discussion}

In practice however, the $9150\times9150\; S$-matrix obtained for $R_0=17\AA$ contains considerably redundant information (for instance, overlap of 2 given orbitals on neighboring sites appears 3528 times) and its large size makes its inversion computationally difficult. Since we are \textit{in-fine} interested in the crystal eigenstates (Bloch functions) that are linear combinations of Bloch sums of L\"owdin orbitals, instead of the procedure sketched above, we construct, orthogonalize and invert a $40\times40\; \Tilde{S}$ overlap matrix between Bloch sums of STOs (at the $\Gamma$ point), truncated to the cut-off distance $R_0$. Although this may be not obvious at first glance, $S$ and $\Tilde{S}$ overlap matrices contain the same "information", in the sense that each matrix element of  $S$ contributes to matrix elements of  $\Tilde{S}$. Once the screening parameters have been determined through iterations of the genetic algorithm (see below), the full S-matrix can be occasionally diagonalized and yields the L\"owdin orbitals. One can then check that orthogonalized Bloch sums of STOs are identical to Bloch sums of orthogonalized STOs. In Figures ~\ref{fig:30}-\ref{fig:31}-\ref{fig:32}, we illustrate the effect of orthogonalization on the different basis functions. It can be seen that the STOs are not too severely modified by the orthogonalization procedure, which indicates that they are a fair zeroth order guess.

As for  the determination of interband matrix elements, the consideration of different bands and different high symmetry points in the Brillouin zone provides more than necessary information for the fit convergence from a computational point of view. However, we observe that somewhat different sets of screening parameters can give similar ``fitness'' parameters. This problem is linked with the fact that upper-band dispersions are necessarily incorrect, due to the non-completeness of finite basis. Hence, optical matrix elements involving s* and d orbitals are poorly represented in k-space, and using them to fit screening parameters can lead to unphysical optimum in the algorithm. Non-completeness is a fundamental difficulty of the EPTB method. On the other hand, the dispersion of physically useful valence and conduction bands is affected by s* and d-band parameters in $\Gamma$, but not significantly affected by their dispersion. In particular, the $d(\Gamma_{12})$ state is known to be a nearly free electron state weakly perturbed by the crystal potential, and its wave function can be accurately determined from empirical pseudo-potential or ab initio calculations. Constraining the parameter space in such a way that the $d (\Gamma_{12})$ orbital agrees with independent empirical pseudo-potential or ab initio calculations cures the problem of under-determination of the set of screening parameters. In table 1, we compare the optimized screening coefficients for GaAs with Slater's atomic screening coefficients \cite{slater30}. Due to the self-content procedure, they obviously depend on the hamiltonian parameters, which are given in Annexe 1for GaAs and Ge. These parameters are a slightly re-worked version of the classical set of ref. Jancu98.

\begin{table}
\centering
\caption{\label{tab:table1}Optimized Slater orbital screening coefficients for gallium (Ga) and arsenic (As), compared with Slater's atomic screening coefficients \cite{slater30}}
\label{tab:table1}
\begin{ruledtabular}  
\begin{tabular*}{0.45\textwidth}{@{\extracolsep{\fill}}lcccc}

Orbital & \multicolumn{2}{c}{Ga} & \multicolumn{2}{c}{As}\\
  & Ref \cite{slater30} & This work& Ref \cite{slater30} & This work \\
\colrule
$4s$ & 1.35 & 1.83 & 1.7 & 1.94\\
$4p$ & 1.35 & 1.77 & 1.7 & 1.79\\
$4d$ & 0.27 & 0.93 & 0.27 & 0.96\\
$5s$ & 0.32 & 1.64 & 0.4 & 1.74\\

\end{tabular*}
\end{ruledtabular}
\end{table}

\begin{table}[b!]
\caption{\label{tab:table2}%
Main interband matrix elements (in eV \AA) for Ge and GaAs at the $\Gamma$ point, calculated from differents models}
\begin{ruledtabular}
\begin{tabular}{clcccc}
& &
\textrm{WF1\footnotemark[1]}&
\textrm{WF2\footnotemark[2]}&
\textrm{LDA\footnotemark[3]}&
\textrm{Hamiltonian\footnotemark[4]}\\
\colrule
\multirow{2}{*}{Ge}   &$P_{0}$ & 7.69 & 10.18& 8.49 & 10.14\\
                      &$Q_{0}$ & 8.29 & 8.42 & 7.32 & 8.70\\
\hline
\multirow{3}{*}{GaAs} &$P_{0}$ & 7.38 & 9.88 & 8.35 & 9.82\\
                      &$P_{1}$ & 0.80 & 0.93 & 1.38 & 0.11  \\
                      &$Q_{0}$ & 8.16 & 8.31 & 7.37 & 8.72\\
\end{tabular}
\end{ruledtabular}
\footnotetext[1]{\footnotemark[2] real space calculation from TB wavefunctions with respectively one and two Slater orbital for each basis element}
\footnotetext[3]{real space calculation from LDA wavefunctions, ABINIT code}
\footnotetext[4]{calculated from Hamiltonian derivation}
\end{table}

We applied the procedure explained above to the prototype systems of Ge and GaAs (Appendix. \ref{Annexe.A}). The electron configuration of Ge is [Ar]$ 3d^{10} 4s^2 4p^2$. In the $spds^*$ model, the deep 3d states are discarded and the basis is formed by the orbitals $4s$, the three orbitals $4p$, the five empty orbitals $4d$ and the empty orbitals $5s$. When building the STO basis $ \mathcal{B} $, we keep fixed the first quantum number $n$ of these orbitals and introduce one adjustable screening parameter {$\alpha $} for each symmetry type. Alternatively, as often done in quantum chemistry, we can improve parametrical flexibility by considering that each element of the starting basis is a linear combination of $q$ STOs instead of one. This does not change much the model, but increases to $4q$ the number of fitted parameters. For GaAs, since there are two different atoms in the unit cell, the number of parameters is twice that for Ge. The fitted screening parameter for Ga and As are given in Table \ref{tab:table1} , and contrasted with the Slater atomic screening constants. Those obtained independently for Ge are close to averaged values for Ga and As.
At the end of the fitting procedure, the global discrepancy on the sum of all interband matrix elements, calculated at the  $\Gamma$, X and L points of the Brillouin zone, is less than $15\%$  with one Slater orbital per atomic state and less than $7\% $ with a linear combination of two Slater orbitals for each atomic state. By changing the relative weights of different spectral or Brillouin zone regions in the genetic algorithm fitness function, the discrepancy  at e.g. $\Gamma$ can be minimized down to the percent range. The residual discrepancy has three distinct physical origins: i)  the difference between orthogonalized STOs and the actual L\"owdin orbitals,  ii) the lack of completeness of the spds* basis, and iii) the missing intra-atomic contribution in the k-space derivation method. 
Table \ref{tab:table2} shows the main interband momentum elements $P_0 \equiv -i\langle s_c|p_{x}|x_v\rangle$, $P_1 \equiv -i\langle s_c|p_{x}|x_c\rangle$ and $Q_0 \equiv -i\langle x_c|p_{y}|z_v\rangle$  obtained for Ge and GaAs at the $\Gamma$ point. Agreement is very good, but a discrepancy observed for the weak matrix element $P_1$, for which real space calculations agree, but differ from the hamiltonian derivation value \cite{Paul}. This might be a trace of the methodological limitation relative to intra-atomic contributions. While further work is required to explore the method limitations and improve the results, the present achievement is already sufficient for most practical purposes.

\begin{figure}[t!]
\includegraphics[width=40mm]{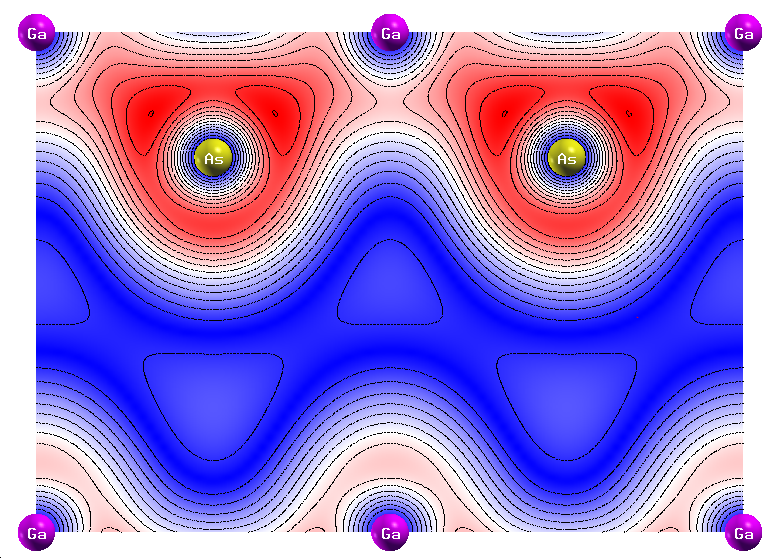} 
\quad
\includegraphics[width=40mm]{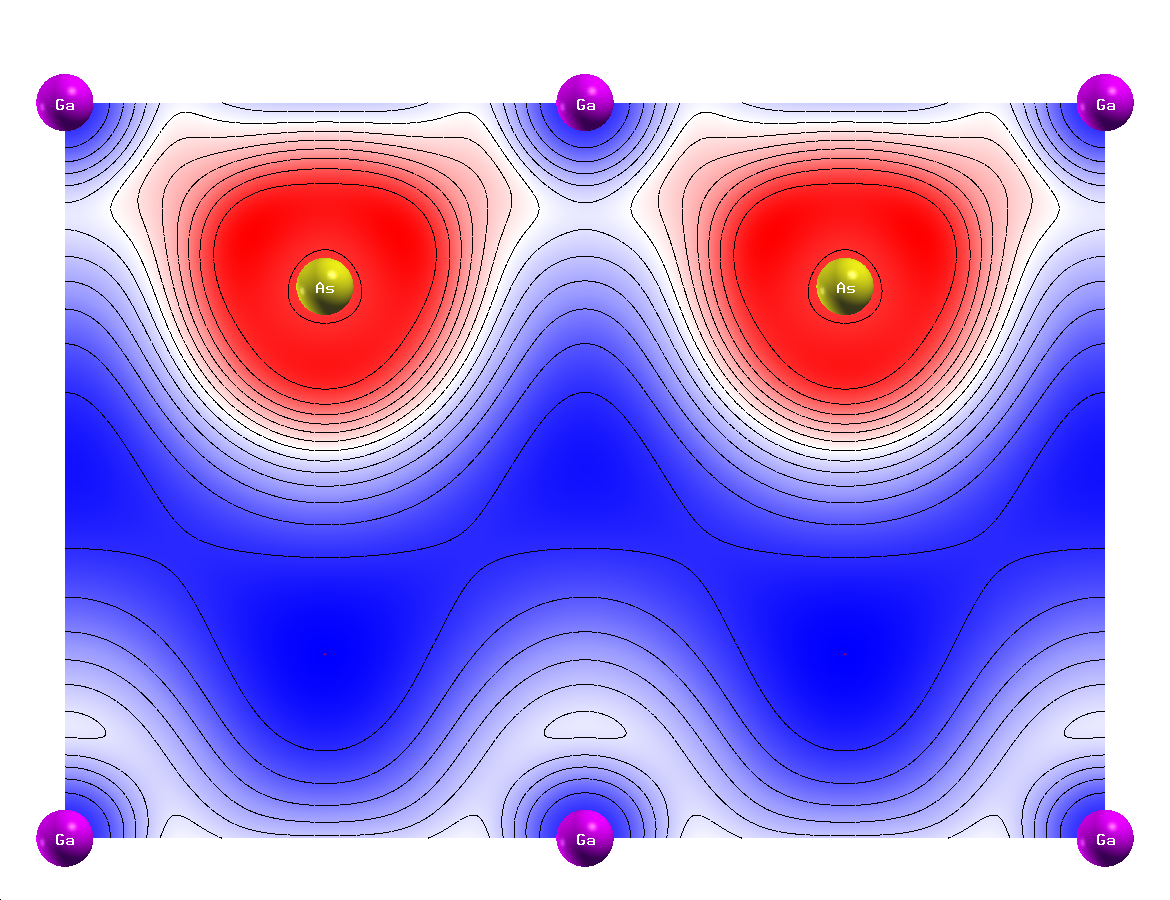}

\includegraphics[width=40mm]{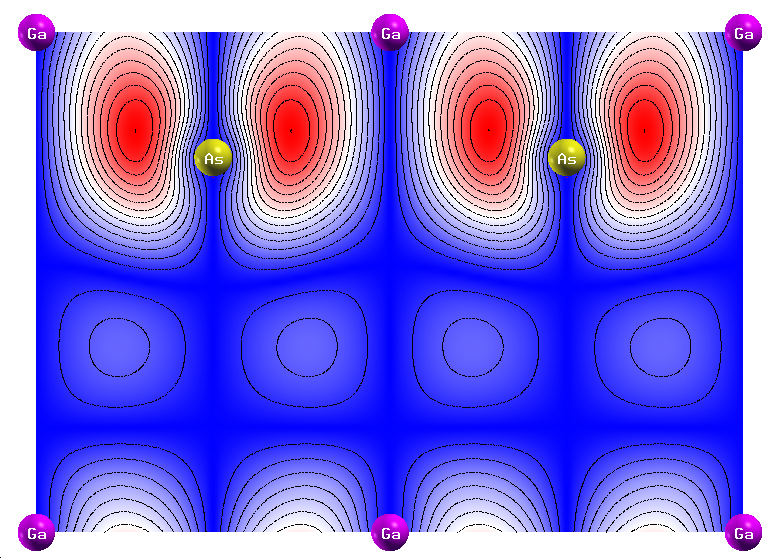} 
\quad
\includegraphics[width=40mm]{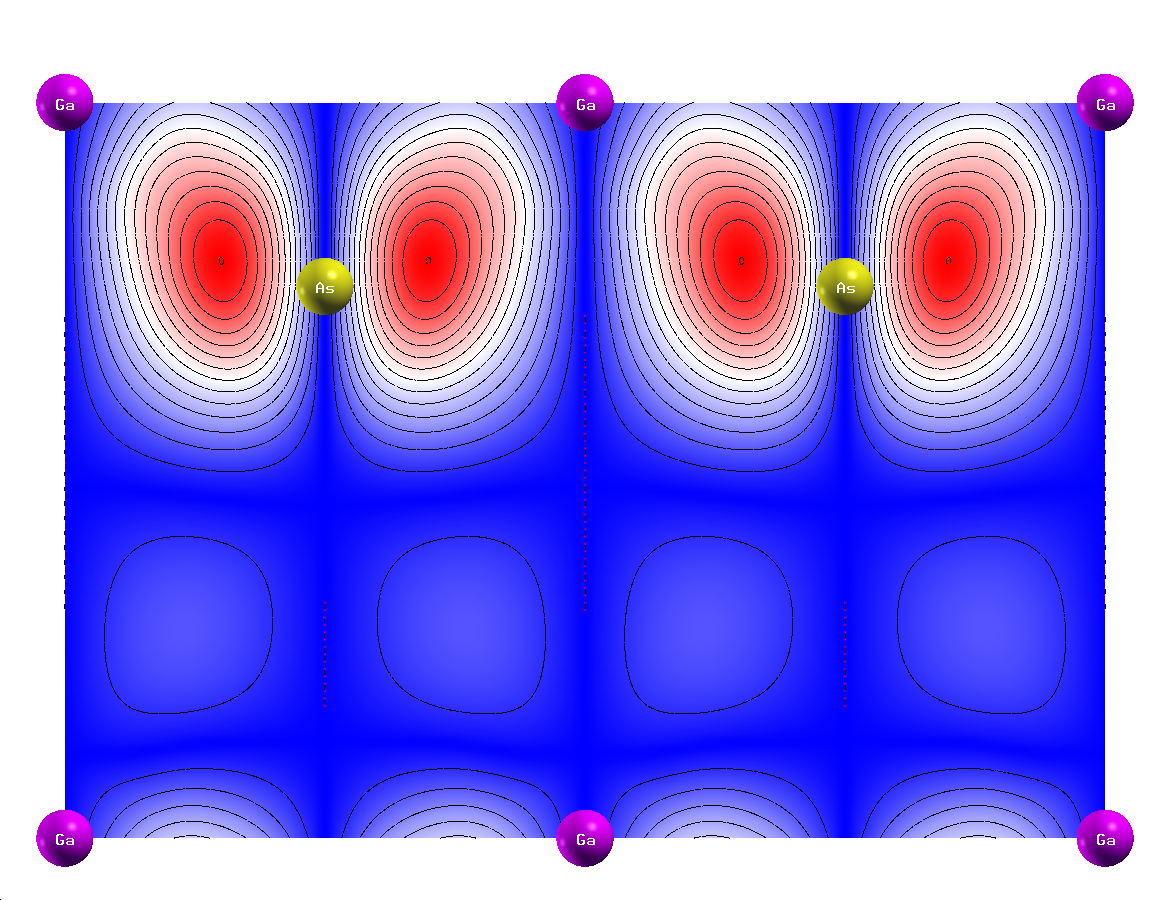}

\caption{\label{fig:fig1} Isodensity contours of the S and Y'= X+Y valence Bloch function at the zone center in bulk GaAs in the plane (110). TB calculation (left) is compared with ABINIT calculations (right). }
\end{figure}

\begin{figure}[t!]
\includegraphics[width=40mm]{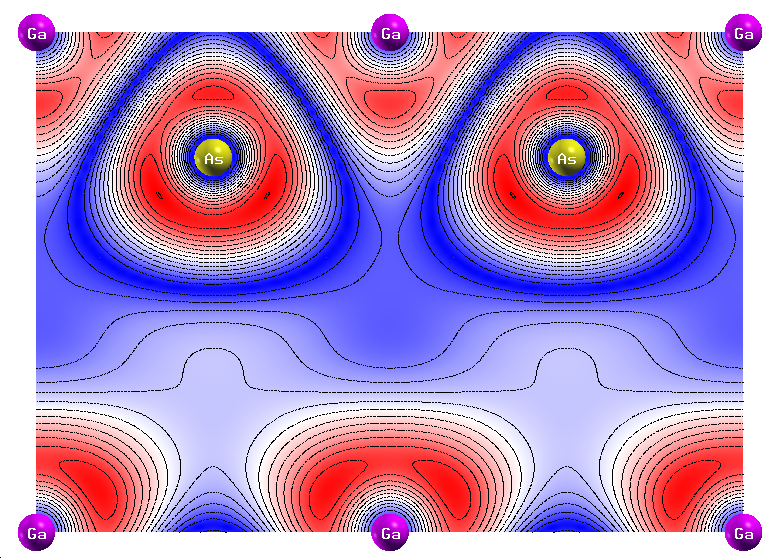}
\quad
\includegraphics[width=40mm]{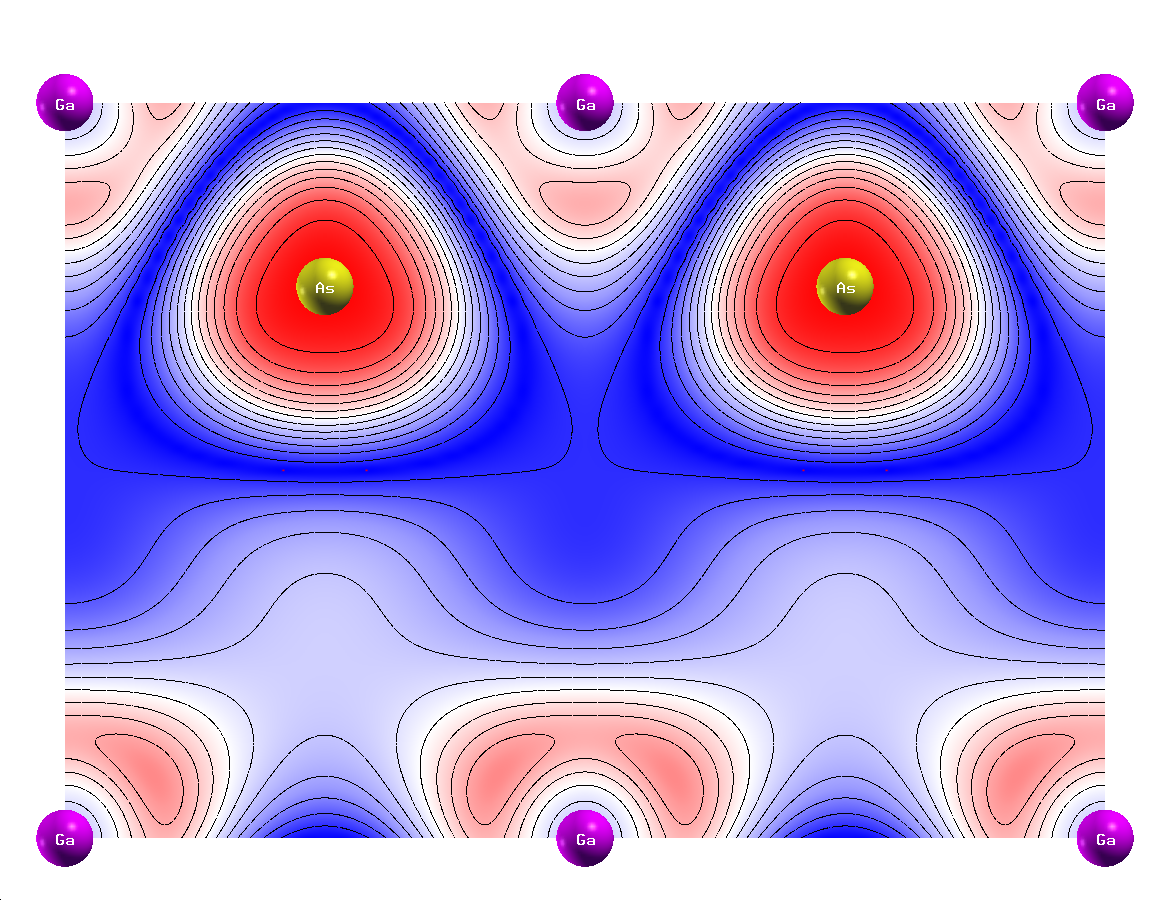}

\includegraphics[width=40mm]{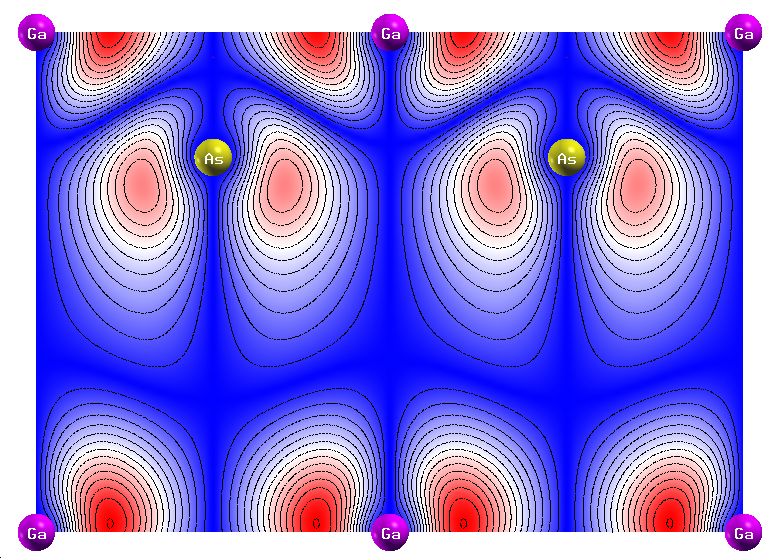}
\quad
\includegraphics[width=40mm]{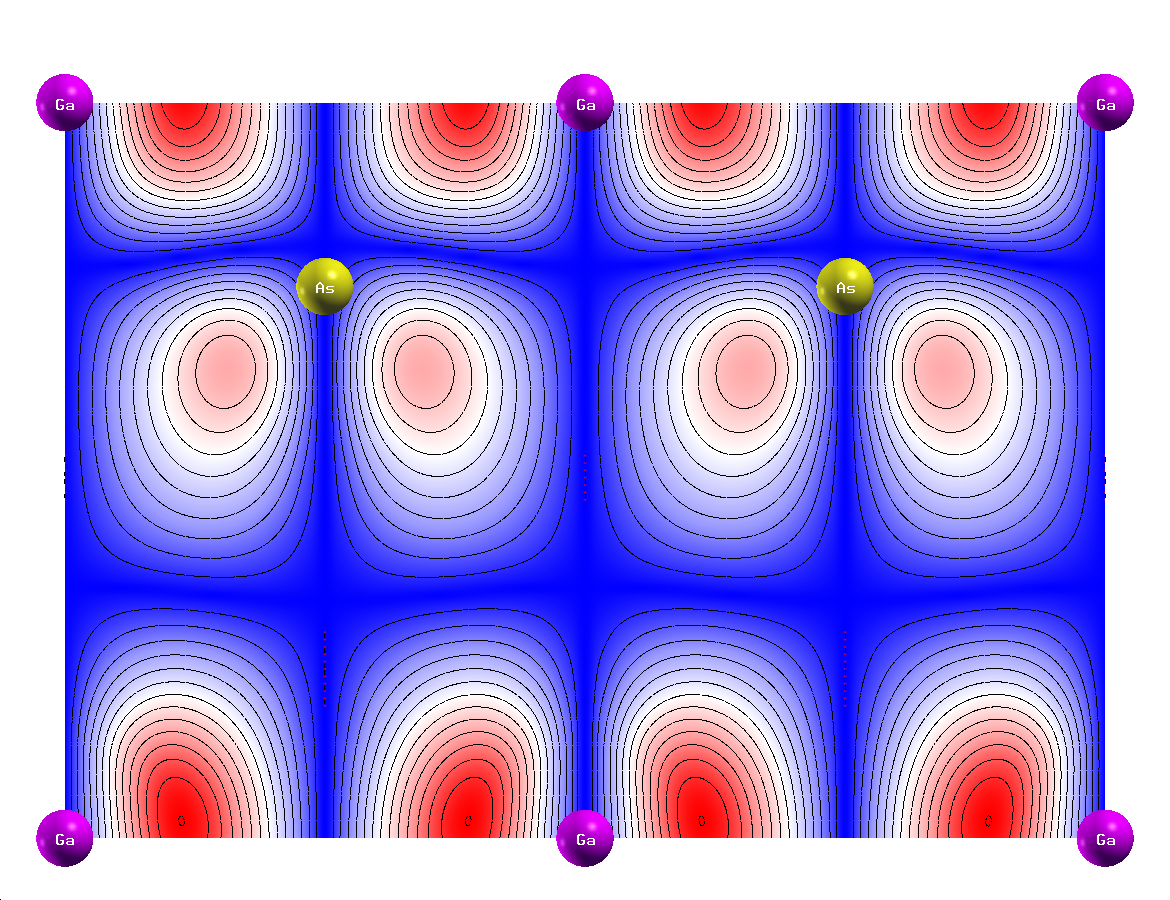}

\includegraphics[width=40mm]{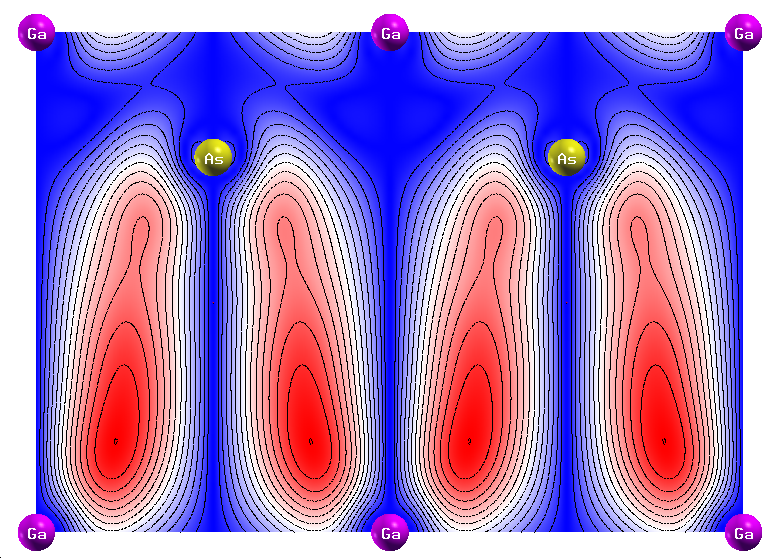}
\quad
\includegraphics[width=40mm]{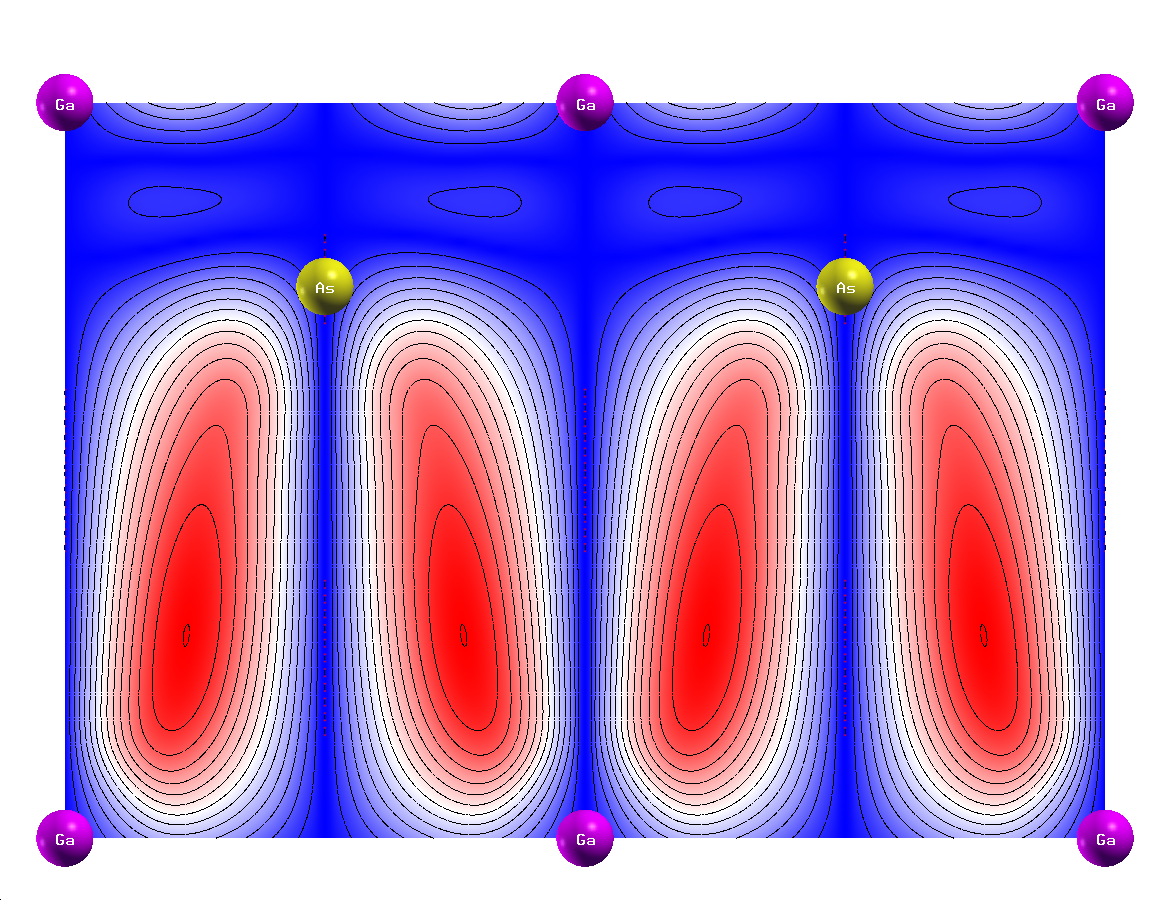}

\caption{\label{fig:fig2} Isodensity contours in the $(110)$ plane for the $S, Y'$ and $D (\Gamma_{12})$ conduction Bloch functions at the zone center in bulk GaAs TB calculation (left) and ABINIT calculations (right)}
\end{figure}

Once screening parameters best reproducing the interband matrix elements are obtained, the different Bloch functions can be plotted and compared with $ab$ $initio$ calculations. The latter were performed using the ABINIT code \cite{Gonze,Gonze1} in the local density approximation (LDA), completed by self-consistent GW correction. Figures ~\ref{fig:fig1}-\ref{fig:fig2} show (110) plane isodensity contours of wavefunctions in bulk GaAs at the Brillouin zone center. Fig.~\ref{fig:fig1} shows valence band states $s_v$ and $y'_v \equiv y_v-x_v$, in both TB and $ab$ $ initio$ calculations. The overall quantitative agreement is very good, since the overlap between TB and $ab$ $initio$ densities is always better than $95\%$. Yet, TB wavefunctions appear somewhat less localized in the sense that they have larger density in regions where the $ab$ $initio$ density is almost zero, and the Ga / As asymmetry is more pronounced in the ABINIT result. The most significant difference is for the deep $s_v$ state near the atomic sites, for which TB density is significantly smaller. This probably reflects the difference of projection basis between the two models: TB wavefunctions are expanded in the basis of Slater orbitals which have a node at the atomic sites while the ABINIT wavefunctions are expanded in a basis of plane-wave functions that may be maximum on the atomic sites. The wavevector cut-off used in the ABINIT calculations is important, because this approach cannot describe the region located less than $1/{k}_{cut-off}$ from atomic sites. Yet, cut-off does not suffice to explain the observed difference. In our TB approach, finite on-site value for the  $s_v$ state results from the contribution of neighboring atoms and is quite sensitive to the STO screening parameters.

In Fig.~\ref{fig:fig2}, we show the conduction Bloch functions also calculated with the same two models. Again TB wavefunctions are very similar to those calculated in the LDA+GW approximation, except for a significant difference for the $s_c$ state density in the vicinity of atomic sites. In order to clarify this issue, we used the SIESTA code, which is based on DFT expanded in a strictly localized orbitals set. SIESTA results for $s_c$ actually agree very well with our TB results (Appendix. \ref{Annexe.B}).

We note that electron hyperfine interaction constants, that are well documented, scale as ${s_c}^2(r=0)$ and could serve as a quantitative test. 
A most striking result is the TB ability to reproduce the wavefunctions of the nearly free electron states $s$* and $d$.

\section{exciton fine structure}
The major interest of having a real space representation of wavefunctions is the ability to study many-body problems. In the following, we  illustrate the potential of our wavefunction derivation method by calculating the exciton binding energy and its fine structure due to electron-hole exchange interaction. Coulomb interaction can be specified by the matrix elements:
\begin{equation}
\begin{array}{c}
\langle \alpha',\textbf{k}'_e;\beta',\textbf{k}'_h|U^{eh}|\alpha,\textbf{k}_e;\beta,\textbf{k}_h\rangle \\
\\
\text{Where} \quad \quad U^{eh}={e^2}/{\kappa |\textbf{r}_{e}-\textbf{r}_{h}|}
\end{array}
\end{equation}
 Here,  $|\alpha,\textbf{k}_e;\beta,\textbf{k}_h\rangle$ is the two-particule excited state, $ \kappa$ is the permittivity, and $\Psi_{\alpha\textbf{k}_e}\equiv\langle\textbf{r}|\alpha,\textbf{k}_e\rangle $ and $\Psi_{\beta\textbf{k}_h}\equiv\langle\textbf{r}|\beta,\textbf{k}_h\rangle $ are the Bloch wave functions in electron and hole representations, respectively \cite{Bir,Goupalov}. The present expansion of Bloch functions as linear combinations of Slater orbitals allows to expand the electron-hole interaction in terms of Coulomb matrix elements between STOs:
\begin{widetext}
\begin{equation}
V_{l_{1}m_{1},l_{2}m_{2},l_{3}m_{3},l_{4}m_{4}}= \langle\phi_{l_{1}m_{1}}(\textbf{r}-\textbf{R}_{1}),\phi_{l_{2}m_{2}}(\textbf{r}-\textbf{R}_{2})|U^{eh}|\phi_{l_{3}m_{3}}(\textbf{r}-\textbf{R}_{3}),\phi_{l_{4}m_{4}}(\textbf{r}-\textbf{R}_{4})\rangle
\end{equation}
\end{widetext}
Restricting the expansion to two-center contributions ($\textbf{R}_{1}=\textbf{R}_{3}$ and $\textbf{R}_{2}=\textbf{R}_{4}$), the evaluation of the integrals can be done quasi analytically using the  expansion of the Coulomb potential in terms of spherical harmonics centered on the same site when $\textbf{R}_{1}=\textbf{R}_{2}$ \cite{Jackson}, and a bipolar expansion when $\textbf{R}_{1}\not= \textbf{R}_{2}$ \cite{Nozawa}. Following \cite{resta,zunger}, we introduce a $r$-dependent dielectric constant such that short-range (on-site) interaction is unscreened, while long range interaction is subject to standard dielectric screening. Then, we solve the Bethe-Salpeter equation (BSE), expressed in terms of calculated electron-hole interactions and energies that are obtained from TB calculation. The BSE is an eigenvalue problem of infinite dimensionality:
\begin{equation}
\begin{array}{l}
(E_{c,\mathbf{k}+\mathbf{Q}/2}-E_{v,\mathbf{k}-\mathbf{Q}/2})A_{vc\mathbf{k}} \\
\\
+\int_{V_{BZ}} d^3k' \sum_{v',c'}\langle vc\mathbf{k}|U^{eh}|v'c'\mathbf{k'}\rangle A_{v'c'\mathbf{k'}} =\Omega_S A_{vc\mathbf{k}} \nonumber
\end{array}
\end{equation}

\begin{figure}
\includegraphics[width=0.45\textwidth]{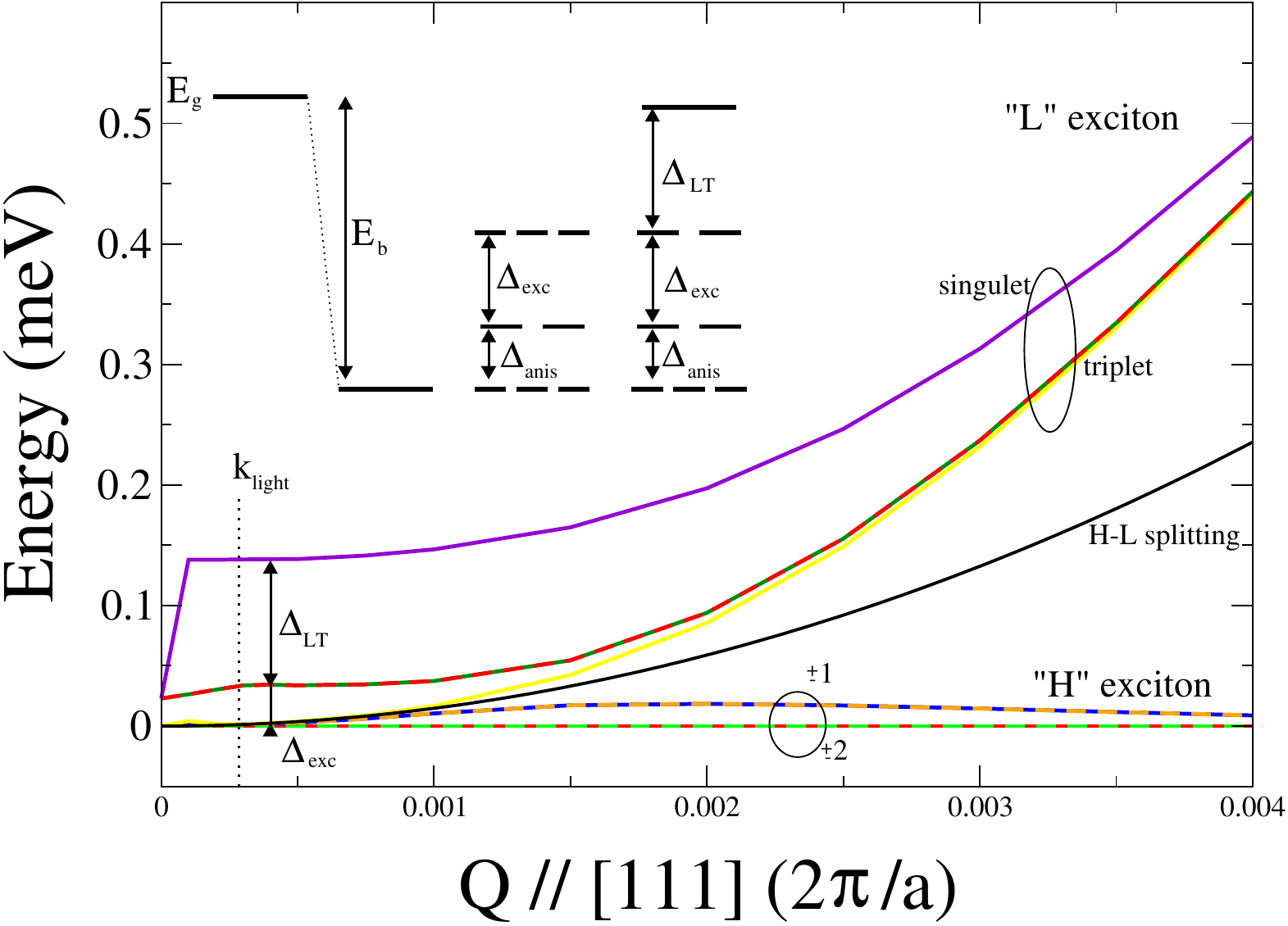} 
\caption{\label{fig:fig6} Dispersion of the exciton states of bulk GaAs for $\mathbf{Q}$ along the [111] direction. The long range exchange is exactly zero at $\mathbf{Q} =0$, and builds up for very small values of $\mathbf{Q}$, for which convergency is more difficult to ensure. The wavevector of light, for which strong polariton features would add to the present picture, is indicated with a vertical line. The inset shows schematically the different contributions to the fine structure of $\Gamma_{8v}\times\Gamma_{6c} $ fundamental exciton.}
\end{figure}

Where $ E_{c,\mathbf{k}+\mathbf{Q}/2}$ and $ E_{v,\mathbf{k}-\mathbf{Q}/2}$ are the electron and hole energies respectively. Resolution of BSE gives the exciton wavefunction components $A_{vc\mathbf{k}} $ and the excitation energies $\Omega_S $. To make the problem tractable continuous integration with respect to $\mathbf{k'} $ was replaced by a discrete scheme. Following \cite{Louie,Louie1}, to calculate the exciton spectra and binding energie at the $\Gamma$ point, the integration was performed over a small region near the position of the band extrema ($|\mathbf{k}|< 0.015$ a.u.). This region was divided into a $11 \times 11 \times 11 $ uniform grid. For an exciton wave vector $\mathbf{Q}=0 $, we find an excitonic binding energy $E_b = 4.75 meV $. In addition, the eightfold degenerate $\Gamma_{8v}\times\Gamma_{6c} $ fundamental excitonic transition is split by short range exchange interaction into one twofold, and two threefold degenerate excitons. The twofold and threefold $J=2$ ``dark excitons'' are split by  $\delta_{anis}=0.02 \mu eV$. This anisotropy splitting is due to the zinc-blend structure which does not allow more than threefold degeneracy. We expectedly find a very small value for $\delta_{anis}$. The $J=1$ ``bright exciton'' threefold state is separated from the $J=2$ states by the short range exchange splitting $\Delta_{exc}$. We get $\Delta_{exc}=20.6 \mu eV$, in agreement with recent experimental determination \cite{PhysRevB.20.3303}. When one moves away slightly from $\mathbf{Q}=0 $, the $J=1$ excitons are further split by the long range exchange interaction into twofold degenerate, optically active transverse excitons, and a longitudinal exciton. The energy difference corresponds to the longitudinal-transverse splitting $\Delta_{LT}$, for which we find a value $\Delta_{LT}=105.3 \mu eV$ in very good agreement with the well documented experimental value. \\

\begin{figure}
\includegraphics[width=0.45\textwidth]{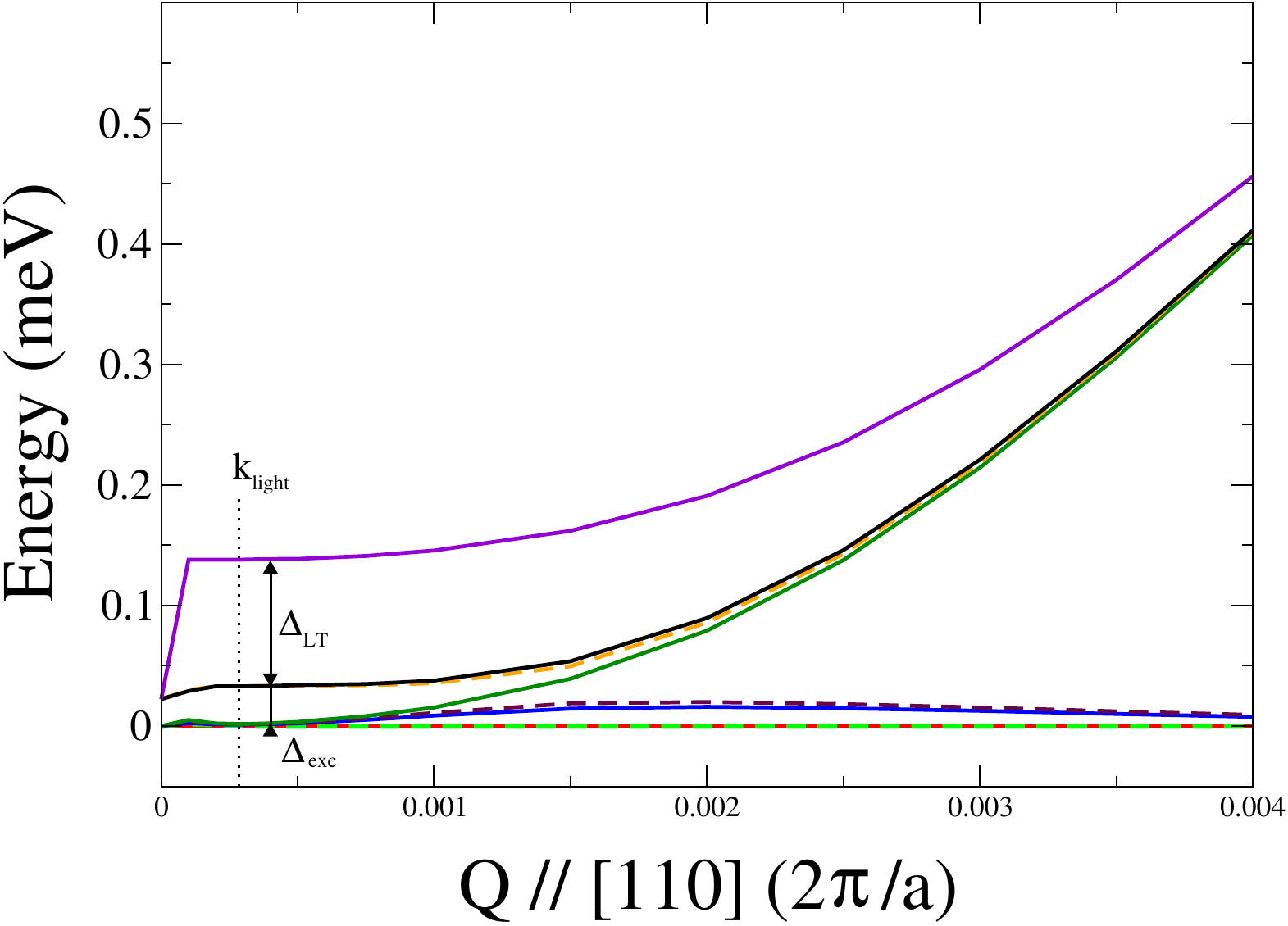} 
\caption{\label{fig:fig7} Dispersion of the exciton states of bulk GaAs for $\mathbf{Q}$ along the [110] direction.}
\end{figure}

Then, we examine the evolution of the exciton fine structure in function of the exciton wavevector $\mathbf{Q}$. Fig.~\ref{fig:fig6} shows the calculated dispersion curves.  For large $\mathbf{Q}$, when the heavy hole light hole splitting becomes larger than exciton binding energy, the exciton splits into a ``heavy'' exciton formed of two twofold degenerate states and a ``light'' exciton formed of one threefold degenerate state and one singlet state. Our calculation shows how energy levels interpolate between the small and large $\mathbf{Q}$ regimes. Finally, we note that when $\mathbf{Q}$ is along the {[110]} direction (Fig.~\ref{fig:fig7}), our results show the full details of exciton state spin splittings, including both contributions of electron and hole spin splittings. Note, however, that linear in k-terms in hole dispersion are not present in current $spds^*$ model.

\section{conclusion}
In conclusion, we have devised a method that allows self-consistent definition of local wavefunctions within the EPTB theory, and successfully used bulk exciton fine-structure as a parameter-free testbed. Extension to nanostructures is straightforward as long as bulk screening parameters are, like other tight binding parameters, transferable to nanostructures. While some fundamental aspects of the method, like the problem of on-site optical matrix elements, still require further clarification, this approach opens a route towards reconciling tight-binding and predictive evaluation of interactions between quasi-particles.

\begin{acknowledgements}
The authors would like to thank Mikhail Nestoklon and Pawel Hawrylak for fruitful discussions.
\end{acknowledgements}

\appendix

\section{hamiltonian parameters of GaAs and Ge}\label{Annexe.A}

\begin{table}[h!]
\caption{\label{tab:tab1}Tight-binding parameters used in calculations.}
\begin{tabular}{|cr|cr|}
\hline 
\multicolumn{4}{c}{Parameters for Ge (eV)}\\
\hline
$              a$ & $    5.6500$&
$          E_{s}$ & $   -3.2967$\\
$        E_{s^*}$ & $   19.1725$&
$          E_{p}$ & $    4.6560$\\
$          E_{d}$ & $   13.0143$&
$       ss\sigma$ & $   -1.5002$\\
$     ss^*\sigma$ & $   -1.9206$&
$   s^*s^*\sigma$ & $   -3.6029$ \\
$       sp\sigma$ & $    2.7985$&
$     s^*p\sigma$ & $    2.8176$\\
$       sd\sigma$ & $   -2.8028$&
$     s^*d\sigma$ & $   -0.6209$\\
$       pp\sigma$ & $    4.2540$&
$          pp\pi$ & $   -1.6510$ \\
$       pd\sigma$ & $   -2.2138$&
$          pd\pi$ & $    1.9001$\\
$       dd\sigma$ & $   -1.2171$&
$          dd\pi$ & $    2.5054$ \\
$       dd\delta$ & $   -2.1389$&

$\Delta/3$ & $    0.12742$       \\
\hline
\end{tabular}
\end{table}

\begin{table}[h!]
\caption{\label{tab:tab1}Tight-binding parameters used in calculations.}
\begin{tabular}{|cr|cr|}

\hline
\multicolumn{4}{c}{Parameters for GaAs (eV)}\\
\hline
$              a$ & $    5.6500$&
$          E_{s}^a$ & $   -5.9820$\\
$        E_{s^*}^a$ & $   19.4477$&
$          E_{s}^c$ & $   -0.3803$\\
$        E_{s^*}^c$ & $   19.4548$&
$          E_{p}^a$ & $    3.3087$\\
$          E_{d}^a$ & $   13.2015$&
$          E_{p}^c$ & $    6.3801$\\
$          E_{d}^c$ & $   13.2055$&
$         ss\sigma$ & $   -1.6874$\\
$   s_as^*_c\sigma$ & $   -1.5212$&
$   s^*_as_c\sigma$ & $   -2.1058$\\
$     s^*s^*\sigma$ & $   -3.7170$&
$     s_ap_c\sigma$ & $    2.8845$\\
$     s_cp_a\sigma$ & $    2.8902$&
$   s^*_ap_c\sigma$ & $    2.5294$\\
$   s^*_cp_a\sigma$ & $    2.3883$&
$     s_ad_c\sigma$ & $   -2.8716$\\
$     s_cd_a\sigma$ & $   -2.2801$&
$   s^*_ad_c\sigma$ & $   -0.6568$\\
$   s^*_cd_a\sigma$ & $   -0.6113$&
$         pp\sigma$ & $    4.4047$\\
$            pp\pi$ & $   -1.4470$&
$     p_ad_c\sigma$ & $   -1.6034$\\
$     p_cd_a\sigma$ & $   -1.6260$&
$        p_ad_c\pi$ & $   1.8422 $\\
$        p_cd_a\pi$ & $    2.1420$&
$         dd\sigma$ & $   -1.0884$\\
$            dd\pi$ & $    2.1560$&
$         dd\delta$ & $   -1.8607$\\
$\Delta_a/3$ & $    0.1745$       &
$\Delta_c/3$ & $    0.0408$       \\
\hline

\end{tabular}
\end{table}

\section{SIESTA versus ABINIT and TB wavefunctions}\label{Annexe.B}

\begin{widetext}

\begin{figure}[h!]
\includegraphics[width=50mm]{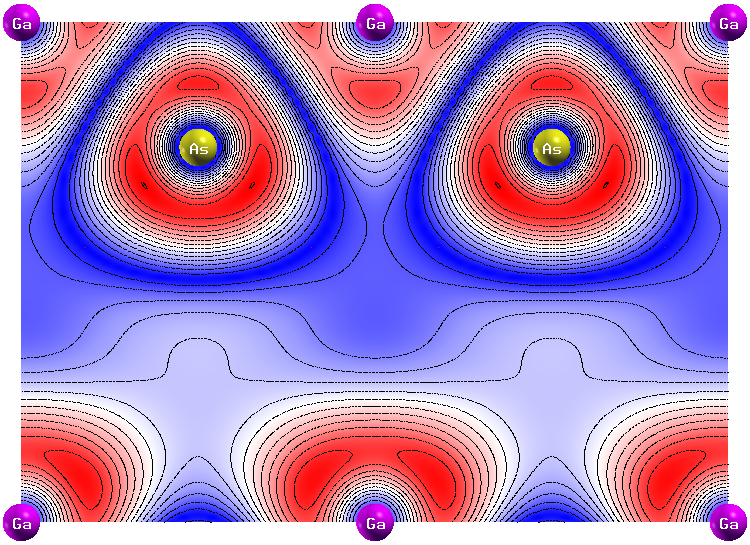}
\quad
\includegraphics[width=50mm]{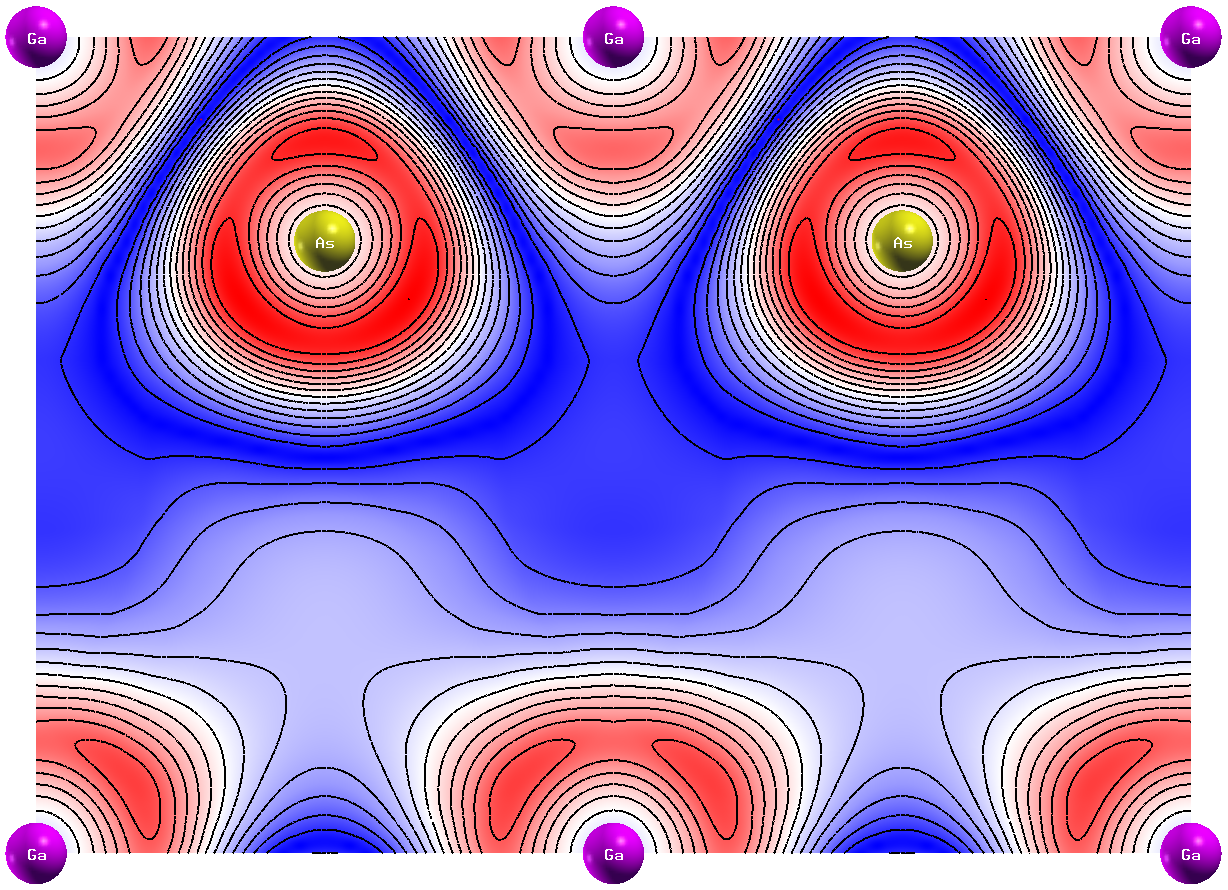}
\quad
\includegraphics[width=50mm]{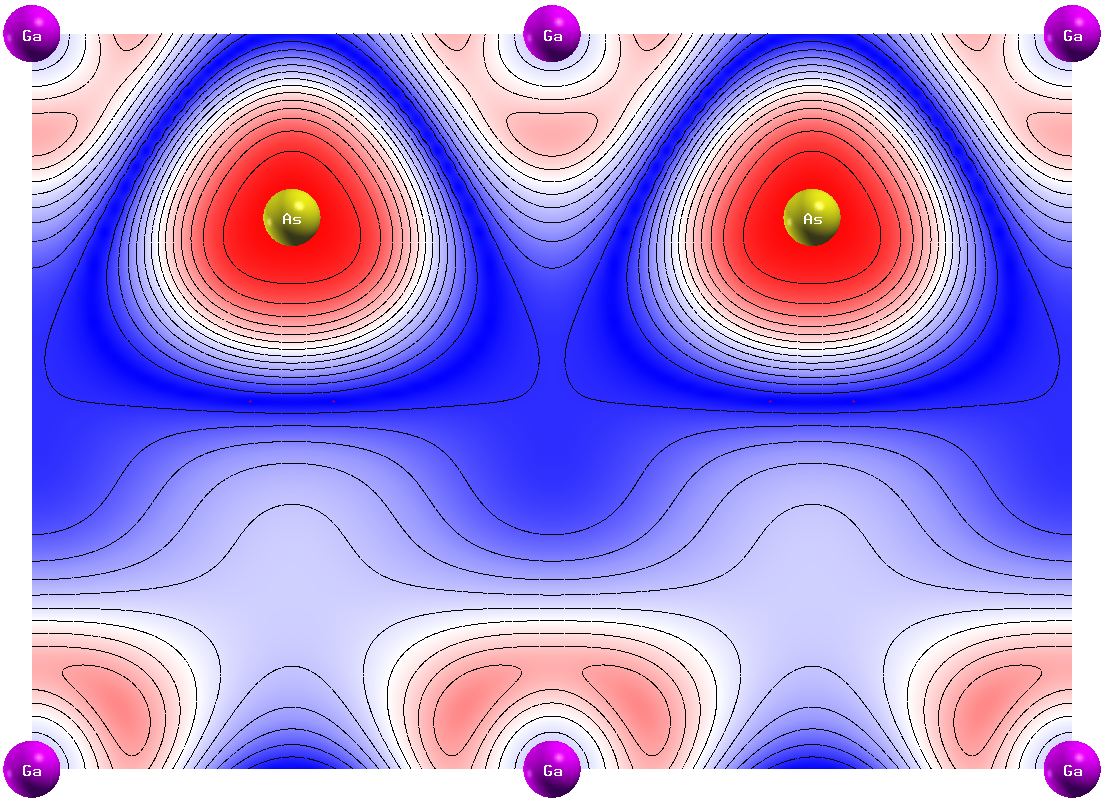}

\caption{\label{fig:fig8} Isodensity contours in the $(110)$ plane for the $s_c$ conduction Bloch function at the zone center in bulk GaAs TB calculation (left), SIESTA calculations (center) and ABINIT calculations (right) }
\end{figure}

\end{widetext}

As discussed in the article, the main differences between the tight-binding wavefunctions and those obtained by ABINIT code was observed in the vicinity of atomic sites. To clarify this issue, we calculated the electronic wavefunction using SIESTA code, which is based in DFT expanded in strictly localized orbitals set. In figure \ref{fig:fig8} we show results for $s_c$ compared to tight-binding and ABINIT calculations. A good agreement between SIESTA calculation and our thight-binding results is observed.

\section{Silicon tight-binding parametrization}

\begin{figure}[h!]

\includegraphics[width=0.45\textwidth]{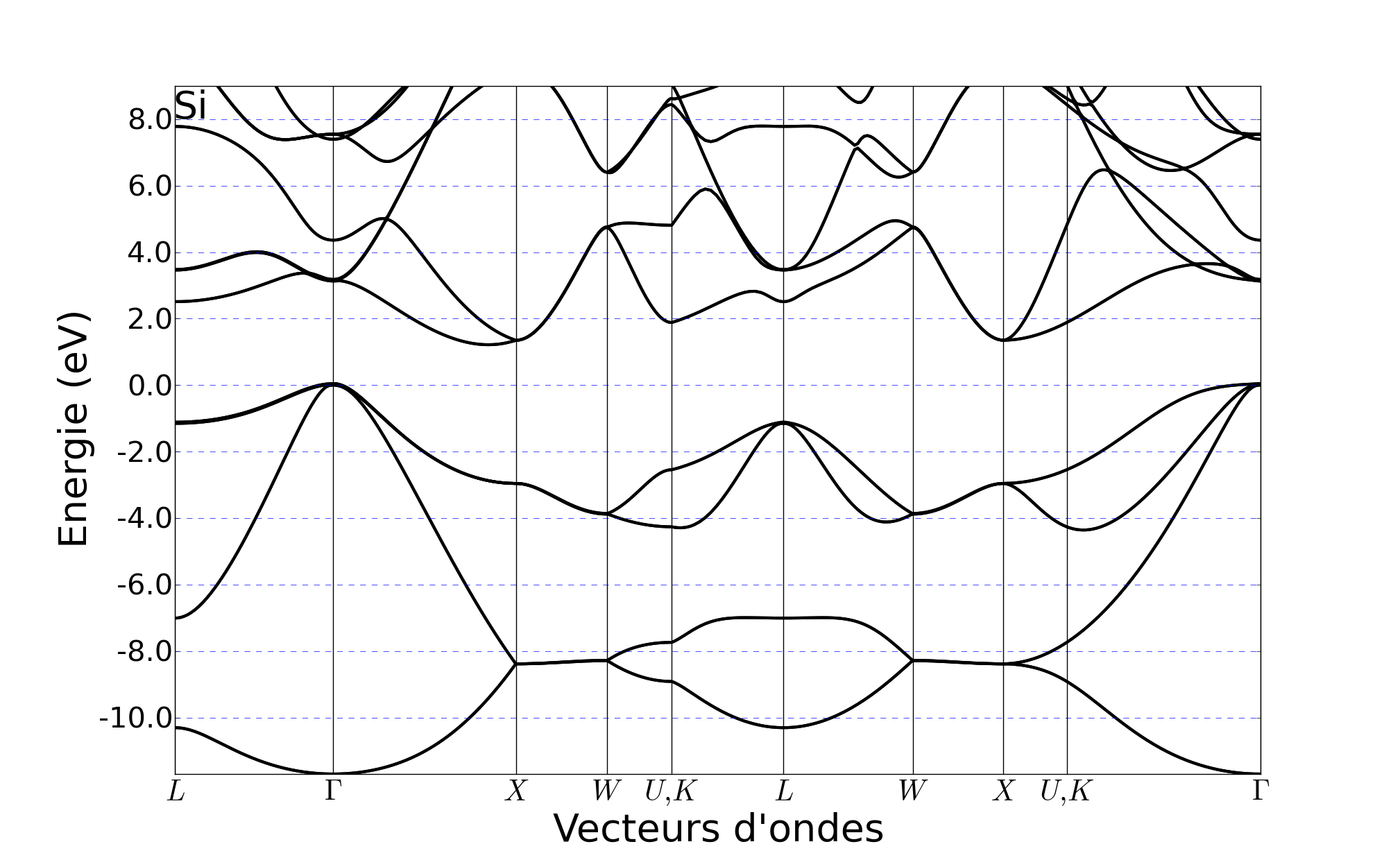}
\caption{\label{fig:fig9} Calculated bulk silicon band structure}
\end{figure}

\begin{figure}[h!]
\includegraphics[width=40mm]{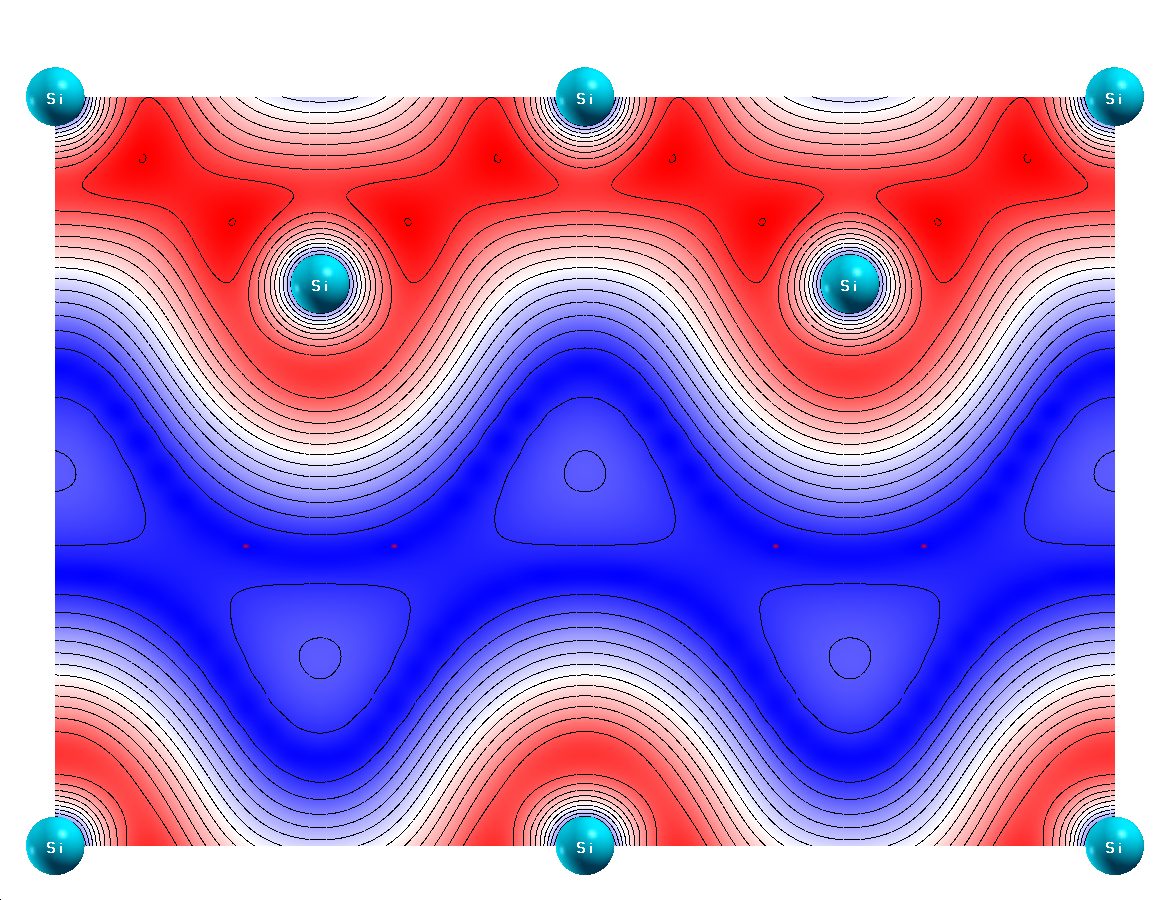}
\quad
\includegraphics[width=40mm]{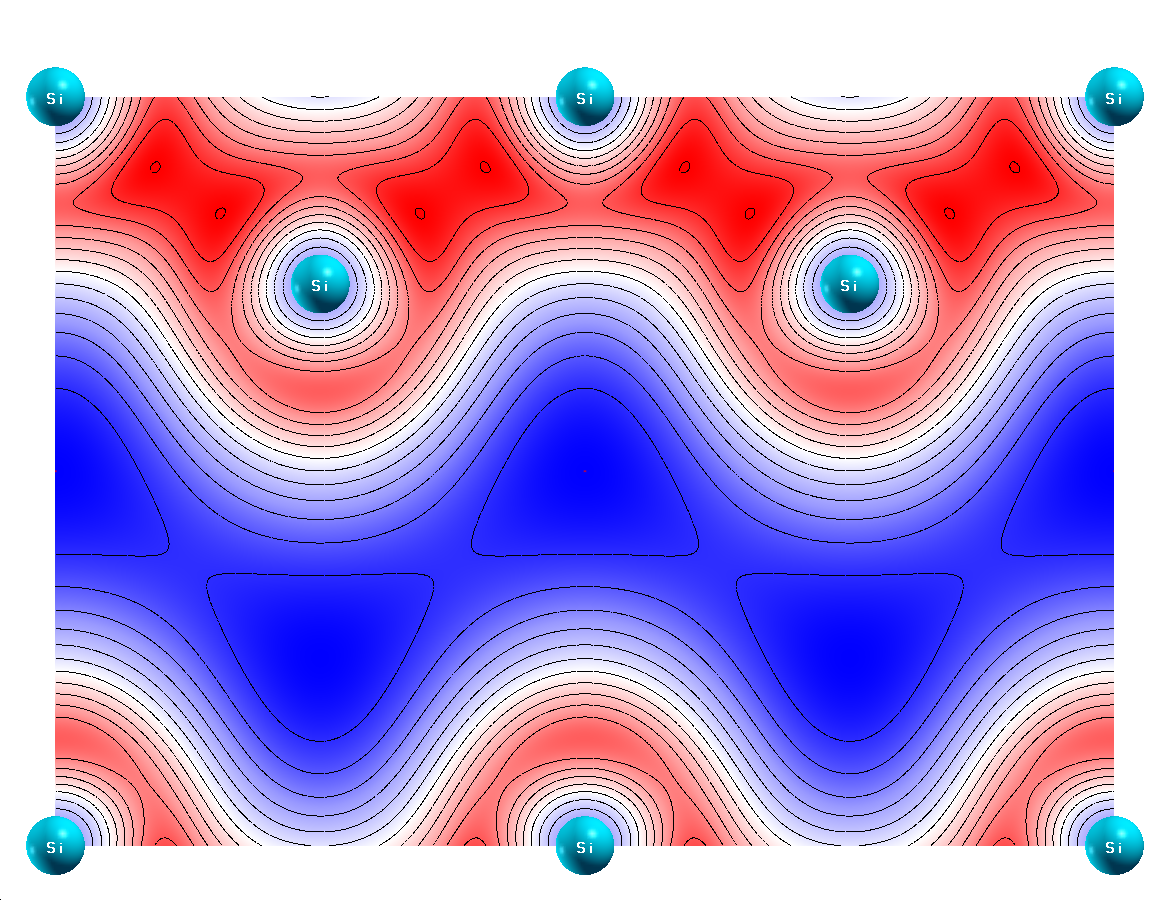}

\includegraphics[width=40mm]{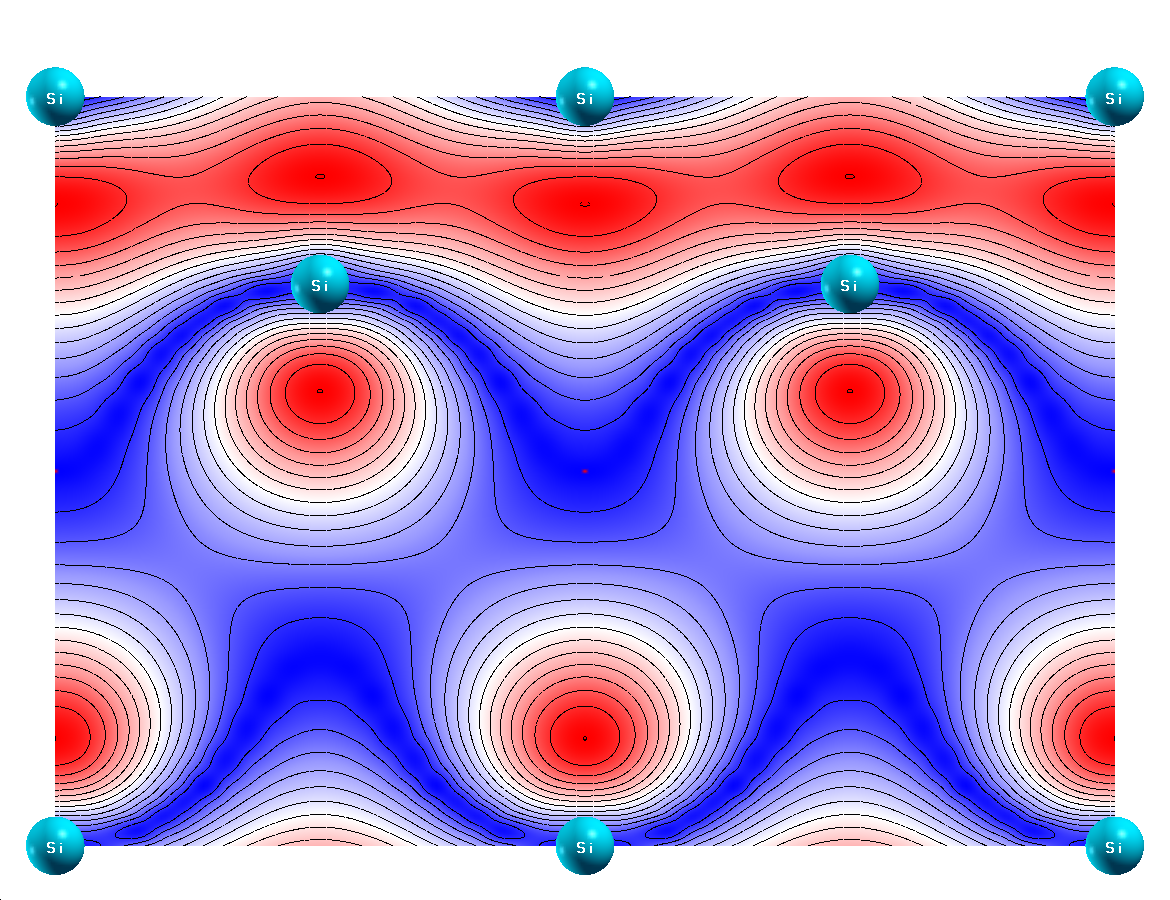}
\quad
\includegraphics[width=40mm]{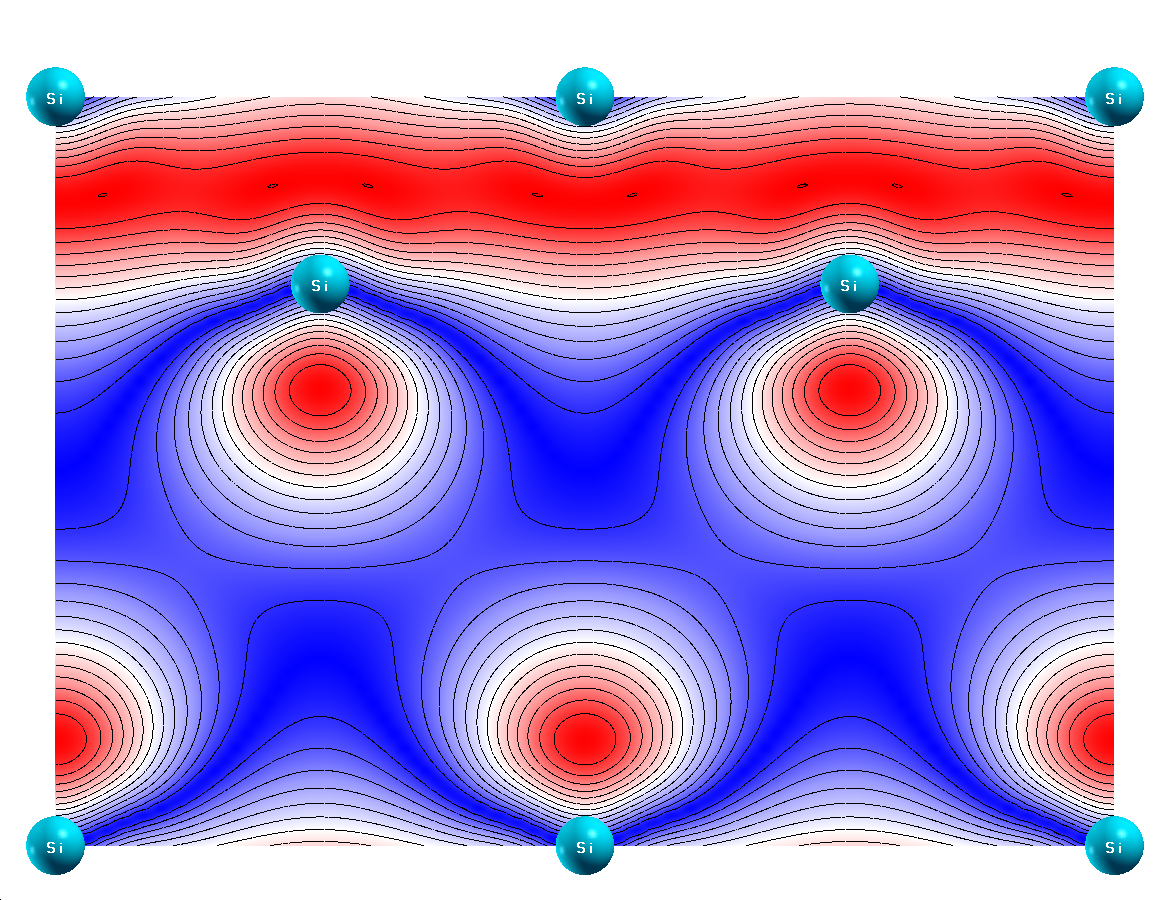}

\includegraphics[width=40mm]{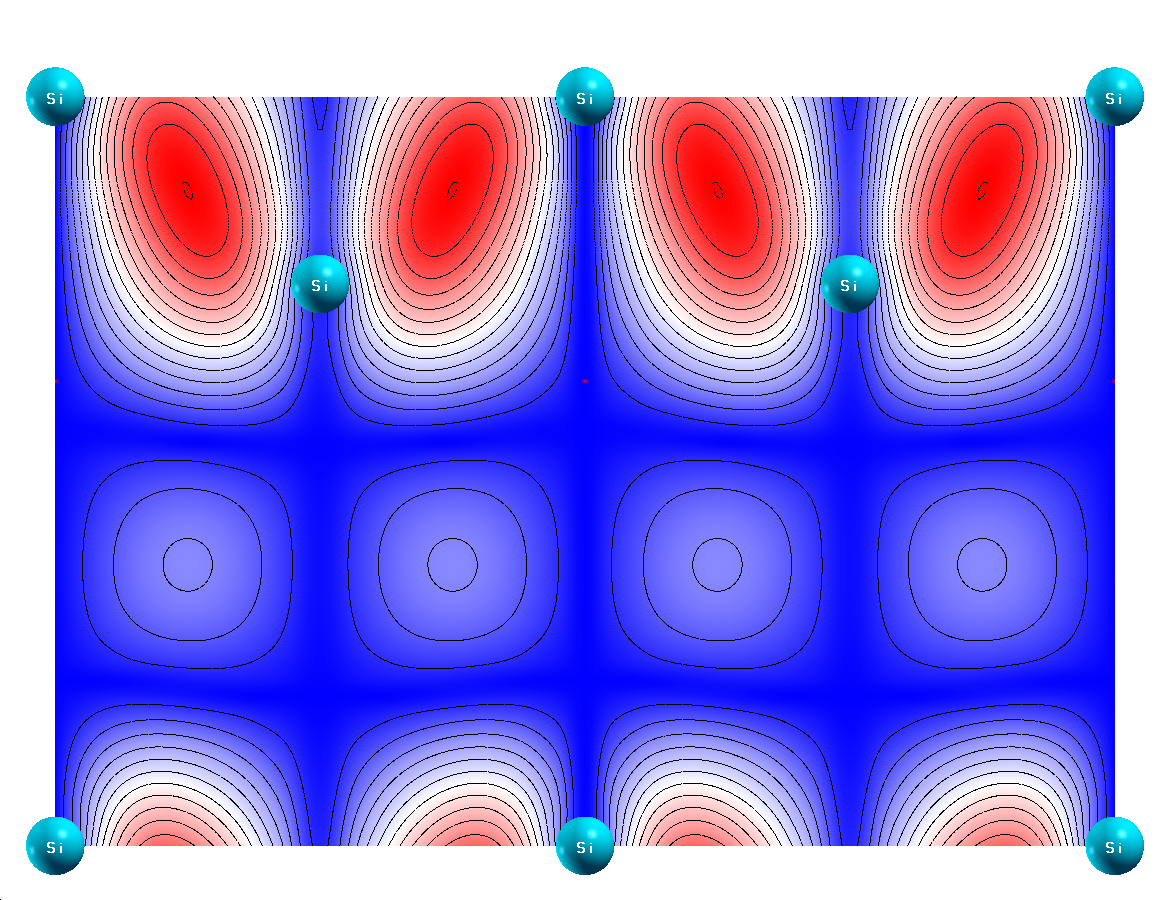}
\quad
\includegraphics[width=40mm]{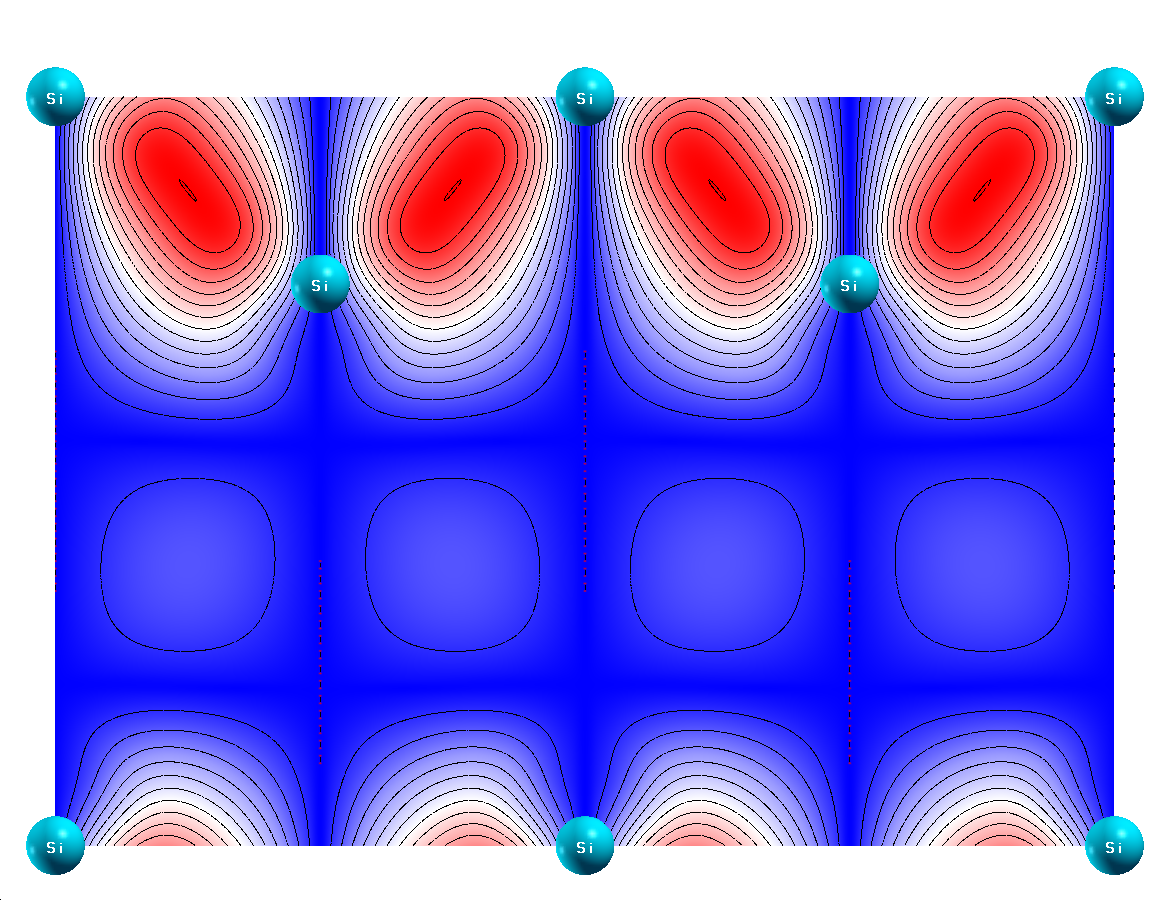}
\caption{\label{fig:fig10} Isodensity contours in the $(110)$ plane for the $s_v$, $z_v$ and $y_v'$ conduction Bloch functions at the zone center in bulk Si TB calculation (left) and ABINIT calculations (right)}
\end{figure}

\begin{figure}[h!]
\includegraphics[width=40mm]{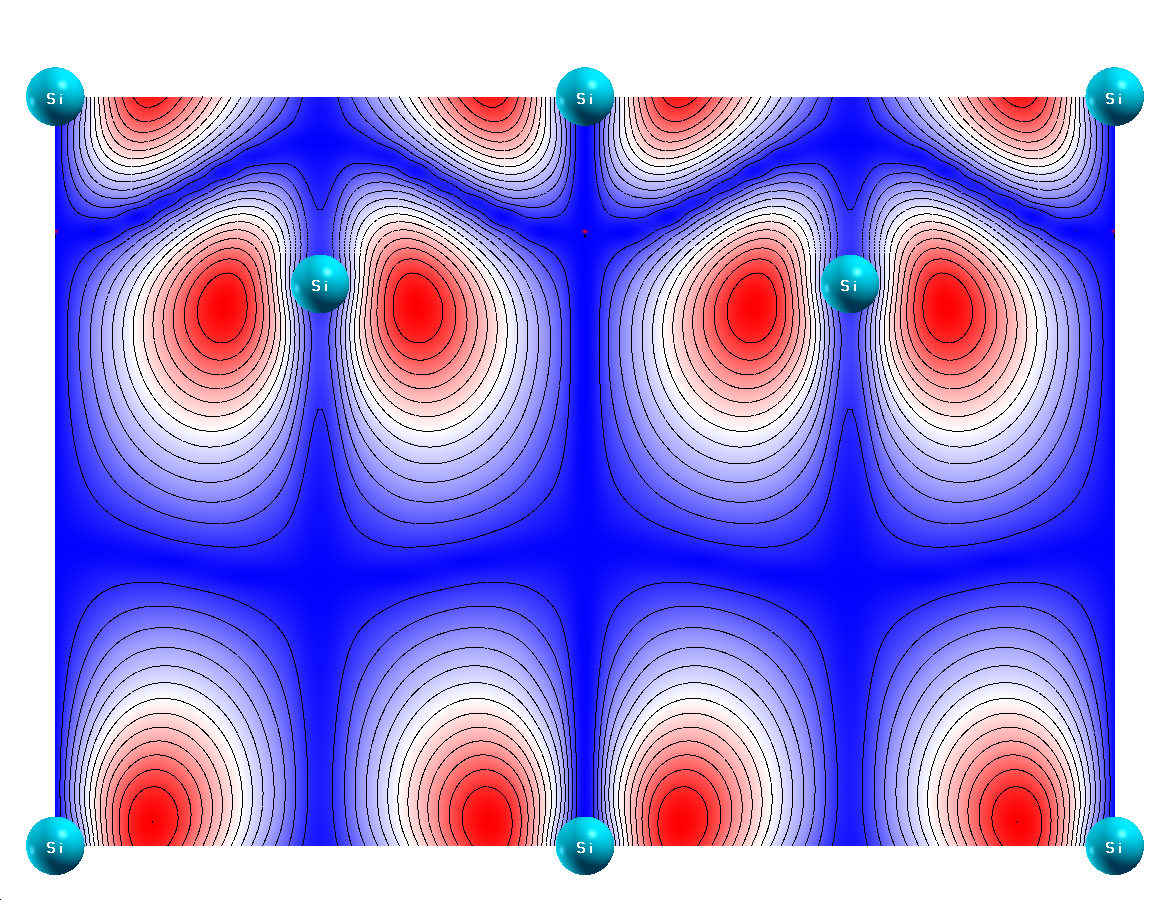}
\quad
\includegraphics[width=40mm]{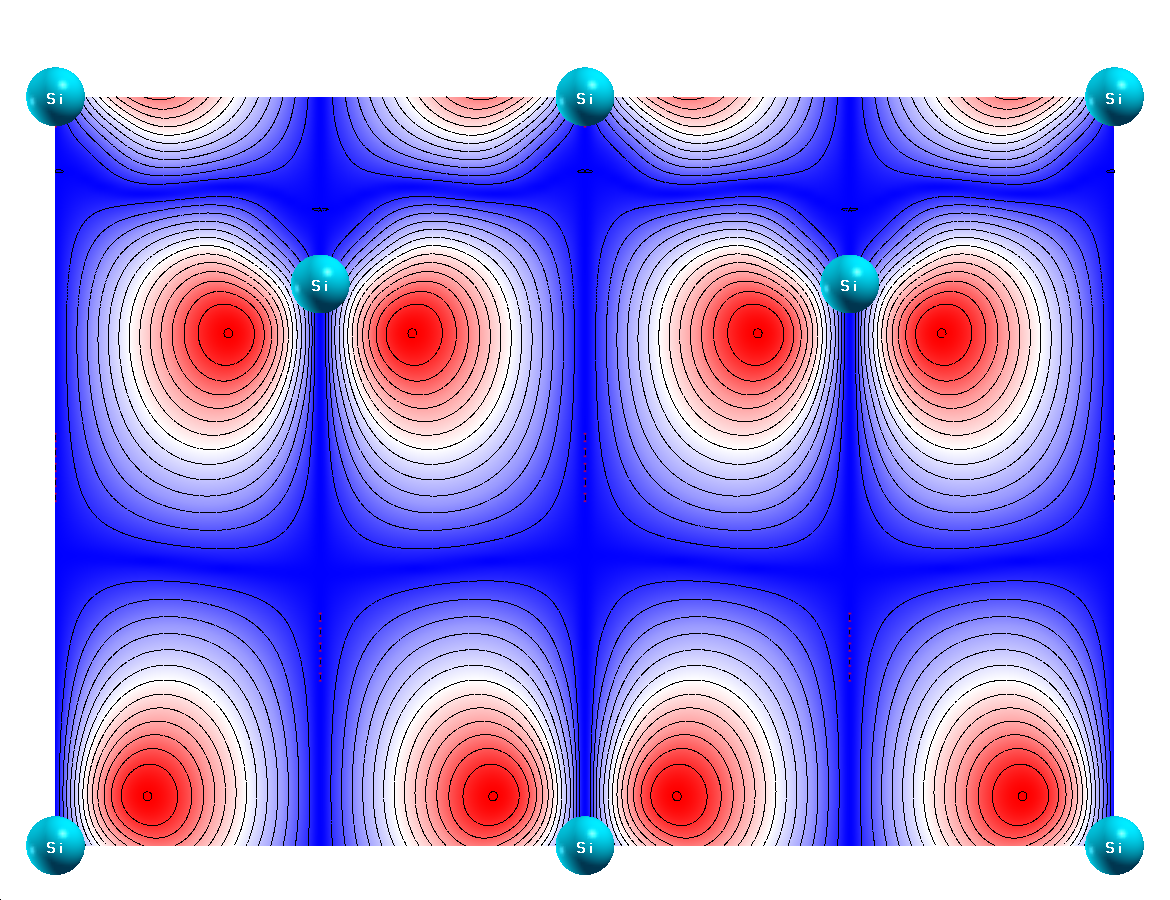}

\includegraphics[width=40mm]{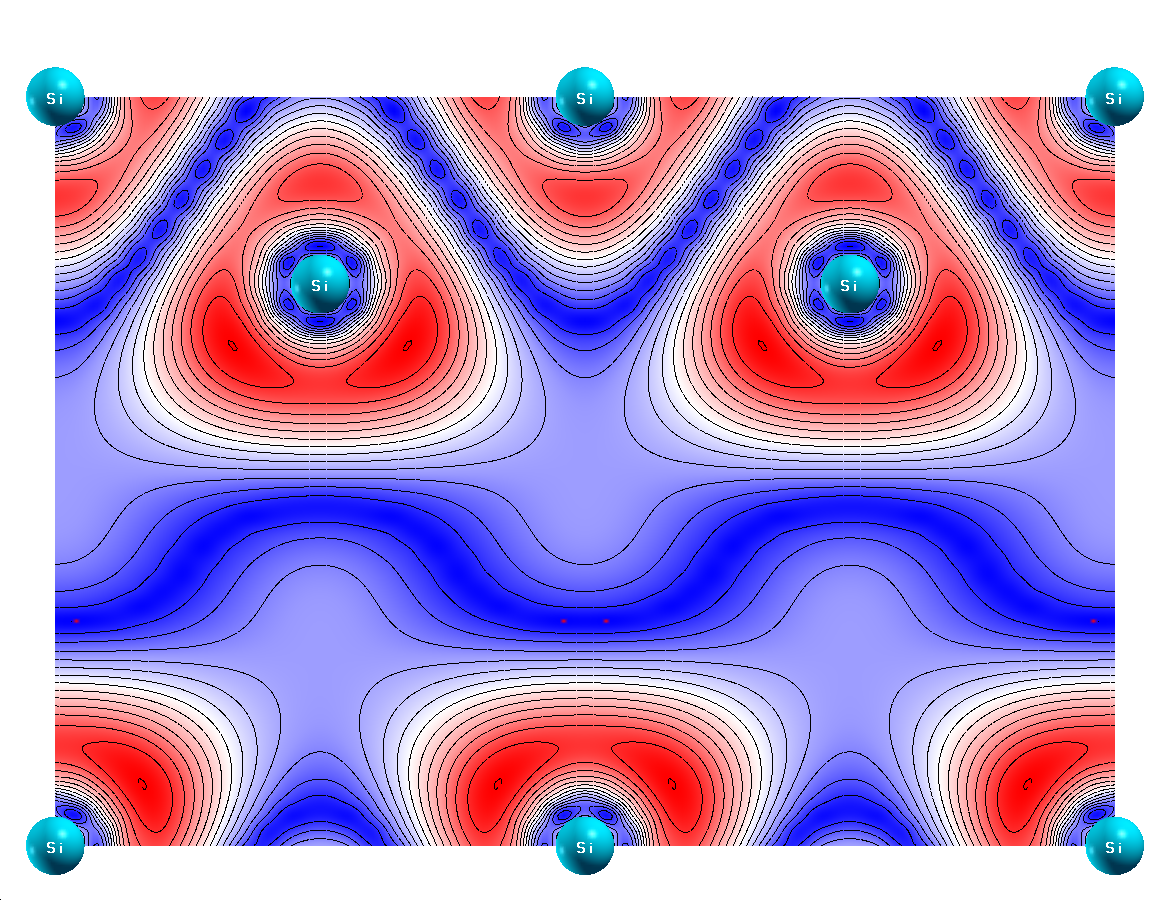}
\quad
\includegraphics[width=40mm]{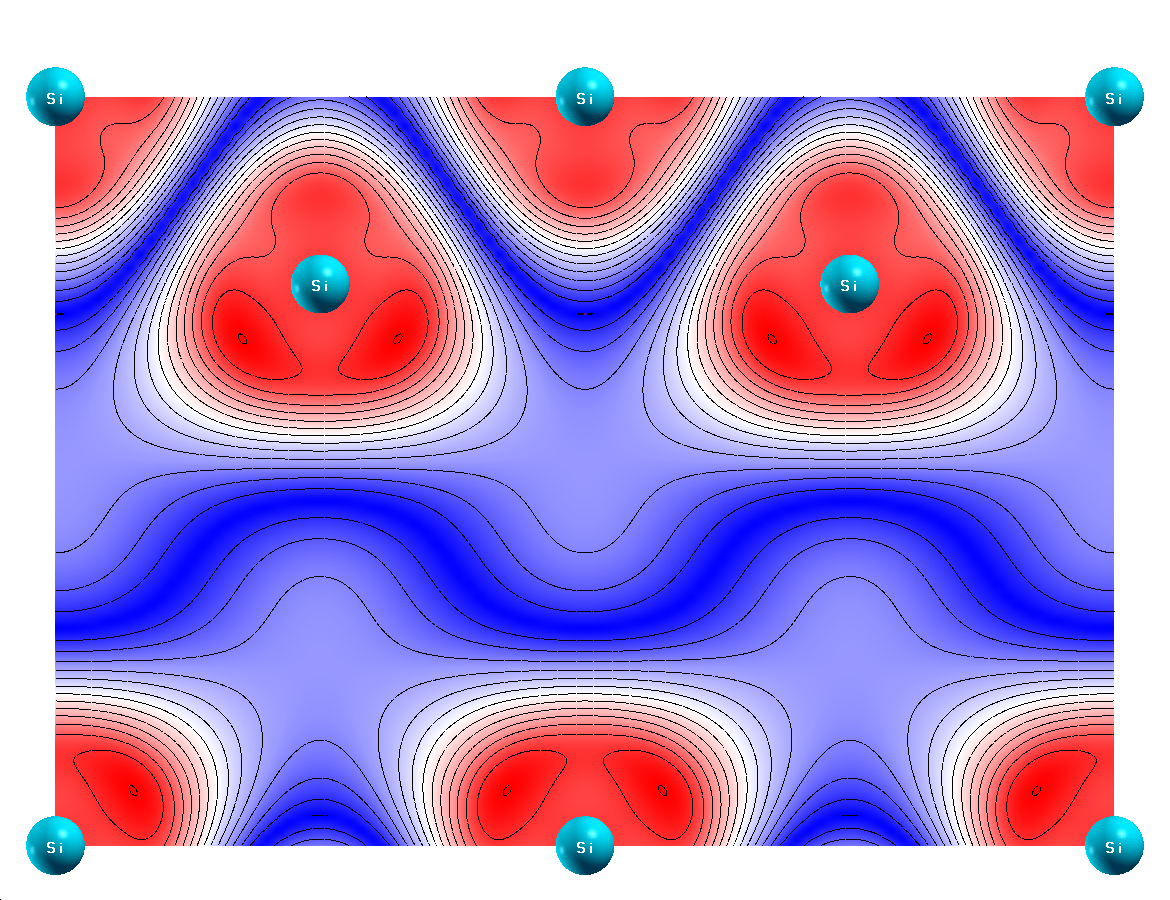}

\includegraphics[width=40mm]{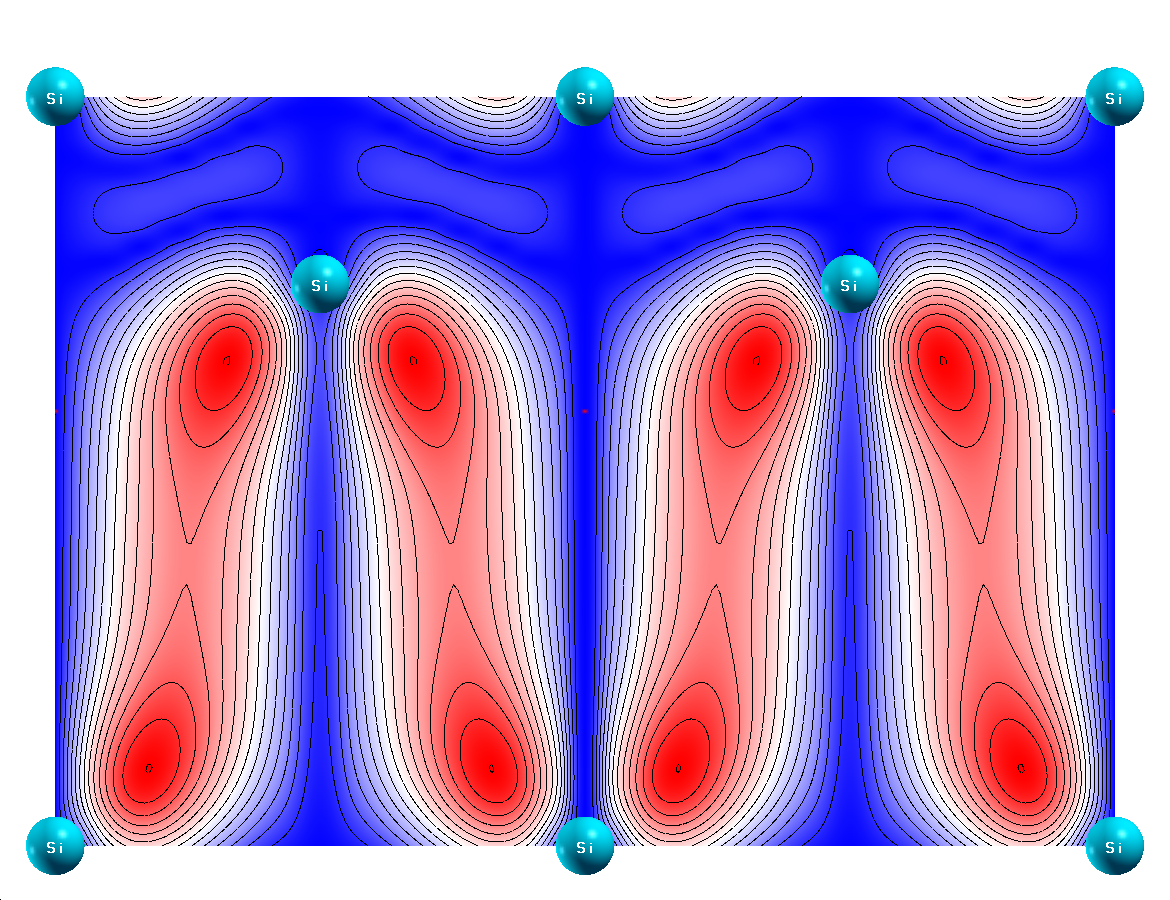}
\quad
\includegraphics[width=40mm]{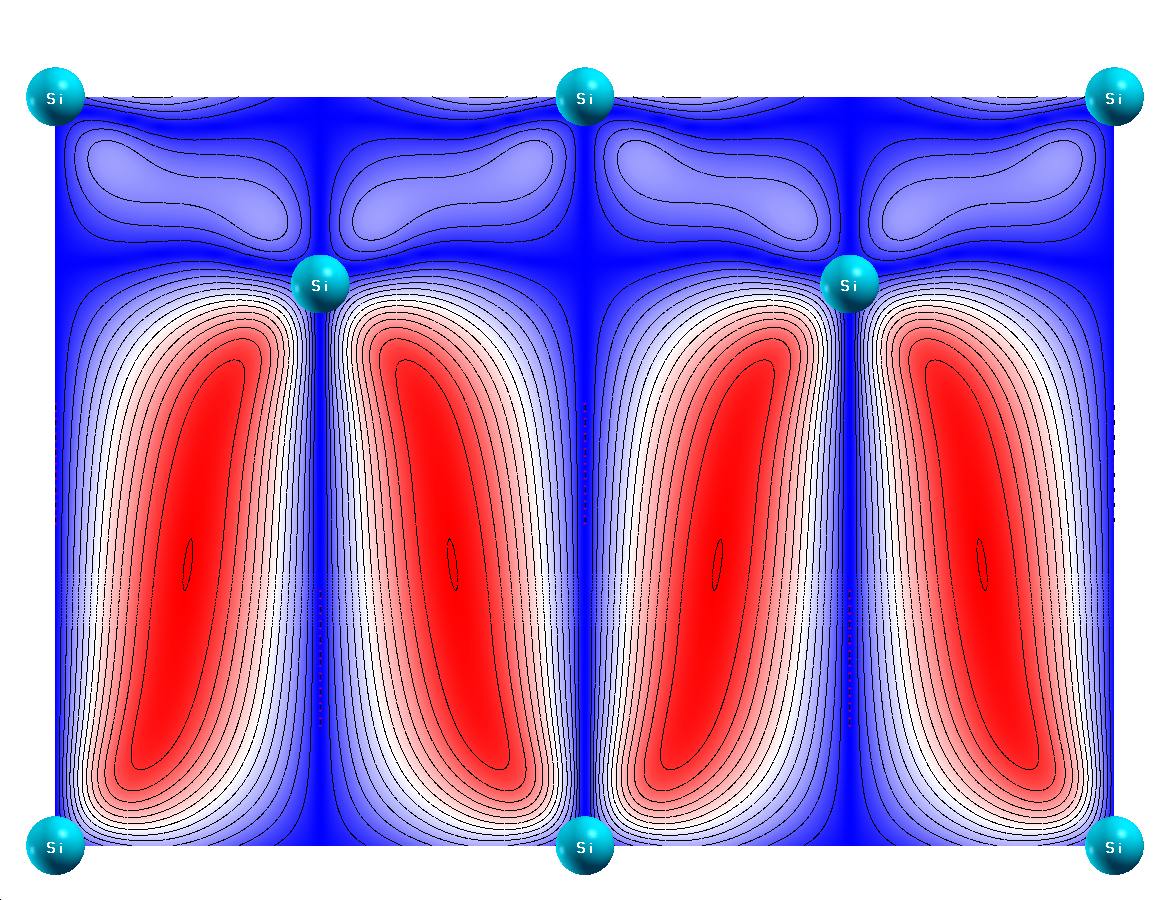}
\caption{\label{fig:fig11}Isodensity contours in the $(110)$ plane for the $y_c'$, $s_c$ and $d (\Gamma_{12})$ conduction Bloch functions at the zone center in bulk Si TB calculation (left) and ABINIT calculations (right) }
\end{figure}

 The efficiency of tight-binding models to calculate electronic properties in silicon has been much discussed\cite{Boykin1} because of its particular electronic band structure presenting an indirect band gap (Figure \ref{fig:fig9})\cite{JM}. A good parametrization of the $sp^3d^5s^*$ model for bulk silicon already exist (Table \ref{tab:tab5}). Table \ref{tab:tab6} show the electronic properties given by this model for the top of valence band and minimum of conduction band. Thess results are in good agreement with experiment. Although the exitonic effects are not relevant in bulk silicon, we applied the procedure described in the article to silicon. Figures \ref{fig:fig10} and \ref{fig:fig11} show our tight-binding wavefunctions compared to ABINIT results.\\

\begin{table}[h!]
\caption{\label{tab:tab5}Tight-binding parameters used in calculations.}
\begin{tabular}{|cr|cr|}
\hline 
\multicolumn{4}{c}{Parameters for Si (eV)}\\
\hline
$              a$ & $    5.4300$&
$          E_{s}$ & $   -2.0386$\\
$        E_{s^*}$ & $   19.9699$&
$          E_{p}$ & $    5.0669$\\
$          E_{d}$ & $   14.8323$&
$       ss\sigma$ & $   -1.8885$\\
$     ss^*\sigma$ & $   -1.5103$&
$   s^*s^*\sigma$ & $   -3.6932$ \\
$       sp\sigma$ & $    2.9607$&
$     s^*p\sigma$ & $    3.5346$\\
$       sd\sigma$ & $   -2.5344$&
$     s^*d\sigma$ & $   -2.0505$\\
$       pp\sigma$ & $    4.3649$&
$          pp\pi$ & $   -1.6285$ \\
$       pd\sigma$ & $   -2.2675$&
$          pd\pi$ & $    2.4736$\\
$       dd\sigma$ & $   -1.5424$&
$          dd\pi$ & $    3.6059$ \\
$       dd\delta$ & $   -1.7157$&

$\Delta/3$ & $    0.0195$       \\
\hline
\end{tabular}
\end{table}

\begin{table}[h!]
\caption{\label{tab:tab6}Some calculated Si band parameters compared 
with available experimental data.}
\begin{tabular}{ccc}
\hline
 & TB & Expt. \\
\hline
$\Delta_{min}$ & $0.85\;X$ & $0.85\;X$ \\
$E_c(\Delta_{min}) $ & $1.17\;eV $ & $1.17\;eV $ \\ 
$m_t(\Delta_{min}) $ & $0.19 $ & $0.19 $ \\ 
$m_l(\Delta_{min}) $ & $0.99 $ & $0.98 $ \\
$\gamma_1$    & $4.5$ & $4.3$ \\
$\gamma_2$    & $0.2$ & $0.3$ \\
$\gamma_3$    & $1.5$ & $1.6$\\
\hline
\end{tabular}
\end{table}

\bibliography{wavefunctions_mai_2014}

\begin{thebibliography}{33}
\bibitem [{TB1()}]{TB1}%
  \BibitemOpen
  \href@noop {} {}\bibinfo {note} {One should however mention that the question
  of on-site transition matrix elements, not included in this approach, is
  still debated within the TB community.}\BibitemShut {Stop}%
\end{thebibliography}
\bibliographystyle{apsrev4-1}

\end{document}